\documentclass[a4paper,11pt]{article}
\usepackage{graphicx}
\usepackage[T1]{fontenc}
\usepackage[utf8]{inputenc}
\usepackage[pdfstartview=FitH,colorlinks=true,linkcolor=blue,anchorcolor=red,citecolor=magenta,urlcolor=blue]{hyperref}
\usepackage[english]{babel}
\usepackage{amsmath,amssymb,titling,authblk}
\usepackage{slashed}
\usepackage{xcolor}
\usepackage{amsfonts}
\usepackage{physics}
\usepackage[margin=3cm]{geometry}
\numberwithin{equation}{section}
\usepackage {color}
\usepackage{url}
\usepackage{cleveref}
\usepackage{hyperref}
\usepackage{cite}
\hypersetup{    colorlinks,  citecolor=blue,   filecolor=blue,    linkcolor=blue,   urlcolor=blue}
\allowdisplaybreaks
\usepackage{mathtools}

\newcommand{\be}{\begin{equation}}
\newcommand{\ee}{\end{equation}}
\newcommand{\bea}{\begin{eqnarray}}
\newcommand{\eea}{\end{eqnarray}}
\newcommand{\eqn}[1]{(\ref{#1})}
\newcommand{\del}{\partial}

\newtheorem{proposition}{Proposition}[section]
\usepackage{mathrsfs}

\numberwithin{equation}{section}
\newcounter{appendice}

\title{\textbf{Bicrossproduct vs.\ twist quantum symmetries in noncommutative geometries: the case of $\varrho$-Minkowski}}
\date{}

\author[1,2]{Giuseppe Fabiano\thanks{giuseppe.fabiano@unina.it}}
\author[1,2]{Giulia Gubitosi\thanks{giulia.gubitosi@na.infn.it, giulia.gubitosi@unina.it}}
\author[1,2,3]{Fedele Lizzi\thanks{fedele.lizzi@na.infn.it, fedele.lizzi@unina.it}}
\author[4]{Luca Scala\thanks{339123@uwr.edu.pl, l.scala.1997@gmail.com}}
\author[1,2]{Patrizia Vitale\thanks{patrizia.vitale@na.infn.it, patrizia.vitale@unina.it}}
\affil[1]{\textit{ Dipartimento di Fisica ``Ettore Pancini'', Universit\`{a} di Napoli {\sl Federico~II}, Napoli, Italy}}
\affil[2]{\textit {INFN, Sezione di Napoli, Italy}}
\affil[3]{\textit {Departament de F\'{\i}sica Qu\`antica i Astrof\'{\i}sica and Institut de C\'{\i}encies del Cosmos (ICCUB),
Universitat de Barcelona, Barcelona, Spain}}
\affil[4]{\textit{ University of Wrocław, Faculty of Physics and Astronomy,
Maksa Borna 9, 50-204 Wrocław, Poland}}

\begin{document}
\maketitle
\begin{abstract}\noindent
We discuss the quantum Poincar\'e symmetries  of the  $\varrho$-Minkowski spacetime, a space characterised by an angular form of noncommutativity. We show that it is possible to give them both a bicrossproduct and a Drinfel'd twist structure. We also obtain a new  noncommutative $\star$-product, which is cyclic with respect to the standard integral measure.
 \end{abstract}

\newpage
\tableofcontents
\section{Introduction}\label{intro}

Quantization of gravity is likely to require some form of quantum spacetime, which may be effectively described by a noncommutative geometry. This, in turn, can be defined by noncommutativity between the spacetime coordinates, similar to what happens with the coordinates of phase space in quantum mechanics. 
Such nontrivial commutation relations  are generally  not covariant under the action of  classical relativity  groups. 
Consequently, quantum spacetimes require the introduction of quantum symmetries,  which are realized by  
 quantum groups of isometries. In the present work  we shall refer in particular to Poincar\'e symmetries. 
 
The latter 
can be approached  either in terms of the noncommutative algebra of continuous functions over the Poincaré group, the quantum Hopf algebra $\mathcal{C}_\star(P)$, or in terms of  the  quantum  Hopf algebra $U_\star(\mathfrak{p})$ obtained from  the universal enveloping algebra of the Poincaré Lie algebra  (the subscript indicating the specific deformation under analysis). 
From the general theory of Hopf algebras, it is possible to show that these two structures are Hopf-dual. The first one, owing to its group-like properties, may be  regarded as a ``quantum symmetry group'', while the second one can be regarded as a ``quantum symmetry algebra''.

Quantum Hopf algebras can be built in different ways. 
Two of the most commonly used structures are the  \emph{bicrossproduct} and \emph{twist}.
The bicrossproduct structure~\cite{Majid:1994cy} can be regarded  as a  quantum generalization of the usual semidirect product of classical  groups. According to \cite{Majid:1994cy}  its mathematical structure is that of a Hopf algebra extension.  It may concern   both the algebra of functions on the group $\mathcal{C}_\star(P)$ and the universal enveloping of the Lie algebra $U_\star(\mathfrak{p})$.
Therefore, 
as in the classical case, it is possible to derive the associated noncommutative spacetime in two  ways.  

One approach involves taking the  quotient of  $\mathcal{C}_\star(P)$ with respect to the  Lorentz Hopf algebra $\mathcal{C}(SO(1,3))$ (which we shall see to be undeformed), so to  obtain the (deformed) translations sector 
that is isomorphic to the spacetime,  as in the classical case. The second  method entails identifying the space 
on which $U_\star(\mathfrak{p})$ acts covariantly, as the spacetime.

The twist approach~\cite{drinfeld} relies on an operator-valued map defined in terms of the generators of some Lie algebra of symmetries.  It   may deform  any bilinear map of a given theory which carries a representation of the Lie algebra. 
It  affects  both the algebra and the coalgebra structure of the associated Hopf algebra of symmetries 
and, when applied to the product of functions on the spacetime, it generates  noncommutativity.

The bicrossproduct and the twist structures are usually considered as alternative options for characterizing the quantum symmetries of noncommutative spacetimes. 
Available examples  
generally deal exclusively with one of these two approaches.\footnote{See however~\cite{juric} where the light-like $\kappa$-Minkowski is associated with two quantum groups of symmetries, one is a bicrossproduct, the other is obtained with a twist.} It is nevertheless interesting to look for quantum isometries in both frameworks, as associated with the  same noncommutative spacetime. 

The $\kappa$-Poincaré quantum group in the so-called ``Majid-Ruegg basis'' (sometimes called ``bicrossproduct basis'') is the most famous quantum group with  a bicrossproduct structure \cite{Majid:1994cy}. It  describes the symmetries of the   $\kappa$-Minkowski spacetime algebra \cite{Lukierski:1993wx}. While the $\kappa$-Poincar\'e algebra  was first obtained in \cite{Lukierski:1991pn, Lukierski:1992dt} by contraction \cite{Celeghini:1990bf}, it  was also   obtained later within a twist approach~\cite{kulish, giaquinto, tolstoy, Borowiec, reshetikhin, bu, Govindarajan:2008qa}. However this requires to enlarge the symmetry algebra, so to include at least the generator of Weyl transformations.  

On the other hand, the Moyal algebra is an example of a noncommutative spacetime with deformed Poincar\'e symmetries which can only be obtained by a twist. Indeed, it is possible to check that it is not the quotient of a bicrossproduct structure~\cite{Kosinski:2004qv, Lukierski}. Its quantum symmetry group, the so-called $\theta$-Poincar\'e, was first discussed in~\cite{Oeckl:2000eg, Wess:2003da, Chaichian:2004za, Chaichian:2004yh}.

In this paper we explore the  relation between the twist and bicrossproduct constructions. We provide an example of a noncommutative spacetime, dubbed $\varrho$-Minkowski, 
whose quantum symmetries can be defined in terms of  both structures. 
While this spacetime was originally proposed in the context of twist deformations of the Poincar\'e group~\cite{Lukierski},  we show that its symmetries can be naturally described also in terms of a bicrossproduct quantum group. The deformed Poincar\'e symmetries obtained following the two approaches are isomorphic quantum groups related one to the other by a nonlinear change of generators. We shall indicate them with $U_\varrho(\mathfrak{p})$ and $\mathcal{U}_\varrho(\mathfrak{p})$. They give rise to different $\star$-products, which are both cyclic with respect to the standard integration measure, an important property towards formulating  gauge invariant theories.
This particular model can therefore serve as a bridge between the two approaches,  facilitating a comparison of the relevant structures and their physical interpretation. 

The paper is organised as follows. In Section~\ref{rhospace} 
we introduce the $\varrho$-Minkowski spacetime and shortly describe  the $\kappa$-Minkowski one, which will serve as a guiding example all over the paper. In Section~\ref{twistedUrho} we review the quantum enveloping algebra $U_\varrho(\mathfrak{p})$~\cite{dimi2} describing the twisted quantum symmetries of the $\varrho$-Minkowski spacetime, and recall the related $\star$-product. 

Sections~\ref{bicro} and~\ref{planew} contain the main original results of the paper. In Section~\ref{sec2} we describe $\mathcal{C}_\varrho(P)$, previously introduced   in~\cite{Lizzi:2021dud, LSV}  via a classical $r$-matrix, and we show by direct calculation that it has a standard bicrossproduct structure. We thus analyse the Hopf algebras $U_\varrho(\mathfrak{p})$ and $\mathcal{U}_\varrho(\mathfrak{p})$ which are both Hopf dual to $\mathcal{C}_\varrho(P)$. 
In Section~\ref{sec3} we prove  that $U_\varrho(\mathfrak{p})$ is not of the bicrossproduct kind. We thus   introduce a suitable non-linear change of generators of the Hopf algebra to obtain the quantum enveloping algebra $\mathcal{U}(\mathfrak{p})$, which has   a bicrossproduct structure. We apply  both a constructive method, as well as a more abstract one, proposed by Majid and Ruegg in the $\kappa$-Poincaré context in~\cite{Majid:1994cy}. The  quantum enveloping algebras $U_\varrho(\mathfrak{p})$ and $\mathcal{U}_\varrho(\mathfrak{p})$ are shown in Sec.~\ref{hopfiso} to be isomorphic. They are both Hopf  dual to $\mathcal{C}_\varrho(P)$, but only the second one is bicrossproduct dual. In analogy with the classical and $\kappa$-Minkowski spacetime, we finally re-derive the $\varrho$-Minkowski spacetime as an appropriate quotient  of the dual $\mathcal{U}^*_\varrho(\mathfrak{p})$ with respect to the action of the Lorentz sub-algebra.

In Sec.~\ref{planew}  we realise the  algebra of noncommuting coordinate functions as finite-dimensional operators and construct operator-valued plane waves in terms of which we define a $\star$-product. Depending on the ordering prescription (time-to-the-right or time-symmetric) we obtain compatible coproducts which are either the bicrossproduct or the twisted one. The time-to-the-right ordering is related with the bicrossproduct structure, in analogy with the $\kappa$-Minkowski case. The time-symmetric ordering, related with the twist, does not have a direct analogue in the $\kappa$-Poincar\'e case, where the same ordering yields the 'standard' $\kappa$-Poincar\'e basis, not descending from a twist. 

Finally, we retrieve the twisted star product already found in \cite{DimitrijevicCiric:2018blz} as the one associated with the time-symmetric ordering, and find a  novel $\star$-product,  related with the   bicrossproduct basis. We show that, similarly to the twisted one,  it is cyclic. However, differently from the latter,  it is not closed. This fact could have interesting consequences to be compared with the results of \cite{DimitrijevicCiric:2018blz, dimi2} where the twisted product has been employed in scalar field theory models. 

A section dedicated to summary and discussion and five appendices containing technical details and review material conclude the paper.

\section{The 
\texorpdfstring{$\varrho$}{varrho}-Minkowski spacetime}\label{rhospace}

In this section we introduce the $\varrho$-Minkowski spacetime, after reviewing   the more popular  and well studied $\kappa$-Minkowski one, which we shall use as a guiding example all over the paper.  We work here with a Lorentzian metric $g_{\mu\nu}$ with signature $(+,-,-,-)$. 

\subsection*{The \texorpdfstring{$\kappa$}{kappa}-Minkowski spacetime}
The $\kappa$-Minkowski spacetime is defined by the commutators 
\begin{equation}
    [x^\mu,x^\nu]=i\ell(v^\mu x^\nu - x^\mu v^\nu ),
\end{equation}
where we use a length (rather than energy) deformation parameter $\ell= 1/\kappa$, for homogeneity with the notation used for $\varrho$-Minkowski. The 4-vector $v^\mu$  characterizes physically different models, based on  whether it is timelike, null or spacelike with respect to the metric $g_{\mu\nu}$~\cite{Lizzi:2020tci}. The timelike $\kappa$-Minkowski with commutation relations 
\begin{equation}
\label{kappacomm}
    [x^0,x^i]=i\ell x^i, \qquad [x^i,x^j]=0 , 
\end{equation}
has been under the spotlight of theoretical investigations for several decades (see for example~\cite{Sitarz:1994rh, Kowalski-Glikman:2002eyl, Dimitrijevic:2003wv, Amelino-Camelia:2007yca, Arzano:2009ci, Lizzi:2018qaf, Ballesteros:2021bhh, Arzano:2022ewc}), given its relevance in phenomenological approaches to the quantum gravity problem~\cite{Amelino-Camelia:2008aez}. 
The symmetry group of the noncommutative spacetime~\eqref{kappacomm} is known as $\kappa$-Poincaré~\cite{Lukierski:1991pn, Lukierski:1992dt, Lukierski:1993wxa} and was studied in several works regarding quantum deformations of the Poincaré algebra (see for example~\cite{Kowalski-Glikman:2002iba, Gubitosi:2011hgc, Gubitosi:2019ymi}). As we will show, in the context of the bicrossproduct structure of the deformed Poincaré symmetries the commutation relations~\eqref{kappacomm} were derived as the dual algebra with respect to the deformed translations generators~\cite{Majid:1994cy}.
This accounts for one of the most interesting properties of the bicrossproduct construction: the possibility of defining a procedure similar to that of the classical case, in which the Minkowski spacetime can be viewed as the dual space to the translation sector of the Poincaré algebra.

\subsection*{The \texorpdfstring{$\varrho$}{varrho}-Minkowski spacetime}
\label{rhominkintro}
The $\varrho$-Minkowski spacetime is defined by the commutation relations 
\begin{align}
[x^1,x^0]&=i\varrho x^2\,,\nonumber \\
[x^2,x^0]&=-i\varrho x^1 \,,\label{algebra}
\end{align}
all other commutators being zero. 

This is a sort of angular noncommutativity, in that the time variable acts as the generator of rotations in the $(x_1, x_2)$ plane. From the measurement point of view, when the coordinates are represented as operators on a Hilbert space, its spectrum will therefore be discrete~\cite{Lizzi:2021dud}.

The noncommutative spacetime~\eqn{algebra}, which has the structure of the Euclidean algebra in 2+1 dimensions (the coordinate $x^3$ being central), was first considered  in~\cite{gutt} (also see~\cite{selene}). 
It has been  analyzed within the twist approach  in~\cite{Lukierski}.  This kind of noncommutative spacetime
might be  physically interesting, as argued in~\cite{Amelino-Camelia:2017pne, Ciric:2017rnf, dimi1, dimi2} and phenomenologically relevant in relation with relative locality~\cite{Amelino-Camelia:2011ycm}. Noncommutative field models with such underlying spacetime have been studied in~\cite{DimitrijevicCiric:2018blz, Hersent:2023lqm } and, in the semiclassical approach,  by~\cite{Kurkov:2021kxa} in the context of Poisson gauge theory. The properties of observers and localization problems in this spacetime were studied in~\cite{LSV}.

A similar type of angular noncommutativity, when time is a commutative variable while $x^3$ plays the role of a proper spatial  rotation,  is considered in~\cite{Gubitosi:2021itz} in the context of a double quantization of both spacetime and phase space. In order to distinguish it from the previous one, the latter has been dubbed  $\lambda$-Minkowski spacetime and is characterized by commutation relations analogous to~\eqref{algebra}, with the roles of $x^3$ and $x^0$ exchanged. Keeping this difference in mind, the analysis carried out for $\varrho$-Minkowski in the subsequent sections can be repeated also for the case of $\lambda$-Minkowski, with minor changes.

\section{The  twisted Hopf algebra $U_\varrho(\mathfrak{p})$  }\label{twistedUrho}
Given a Lie algebra, $\mathfrak{g}$, its  universal enveloping algebra, $U(\mathfrak{g})$, may be deformed to a quantum Hopf algebra if a twist operator is available. This is an invertible map $\mathcal{F}\in U(\mathfrak{g})\otimes U(\mathfrak{g})$ (called Drinfel'd twist), 
which satisfies cocycle and  normalization conditions. We shortly review in App.~\ref{appA} some of its properties. When an admissible twist is available, it is possible to define both a quantum spacetime and its quantum symmetry group.     
            
The Drinfel'd twist for the $\varrho$-Minkowski case was first  introduced in~\cite{Lukierski}:
\begin{equation}\label{rhotwist}
\mathcal{F}=e^{\frac{i\varrho}{2}[P_0\wedge M_{12}]}\,,
\end{equation}
with $P_\mu$, $M_{\mu\nu}$   translations and Lorentz generators of the Poincar\'e algebra $\mathfrak{p}$, respectively.
The corresponding quantum Hopf algebra, $U(\mathfrak{p})$,  was first introduced in~\cite{Ciric:2017rnf}. 
The twisted coproducts of the Lie algebra generators, which we shall need in the forthcoming  sections, read
\be\label{twistedrhoqg}
\begin{array}{r@{}l}
\Delta_\mathcal{F}P_0 =&P_0\otimes 1+1\otimes P_0\,,\\
\Delta_\mathcal{F}P_1 =&P_1\otimes \cos \left(\frac{\varrho}{2}P_0 \right)+\cos \left(\frac{\varrho}{2}P_0 \right)\otimes P_1+P_2\otimes \sin \left(\frac{\varrho}{2}P_0 \right)
-\sin \left(\frac{\varrho}{2}P_0 \right)\otimes P_2\,,\\
\Delta_\mathcal{F}P_2 =&P_2\otimes \cos \left(\frac{\varrho}{2}P_0 \right)+\cos \left(\frac{\varrho}{2}P_0 \right)\otimes P_2-P_1\otimes \sin \left(\frac{\varrho}{2}P_0 \right)
+\sin \left(\frac{\varrho}{2}P_0 \right)\otimes P_1\,,\\
\Delta_\mathcal{F}P_3 =&P_3\otimes 1+1\otimes P_3\,,\\
\Delta_\mathcal{F}M_{12} =&M_{12}\otimes 1+1\otimes M_{12}\,,\\
\Delta_\mathcal{F}M_{13} =&M_{13}\otimes\cos \left(\frac{\varrho}{2}P_0 \right)+\cos \left(\frac{\varrho}{2}P_0 \right)\otimes M_{13}-M_{32}\otimes\sin \left(\frac{\varrho}{2}P_0 \right)
+\sin \left(\frac{\varrho}{2}P_0 \right)\otimes M_{32}\,,\\
\Delta_\mathcal{F}M_{32} =&M_{32}\otimes\cos \left(\frac{\varrho}{2}P_0 \right)+\cos \left(\frac{\varrho}{2}P_0 \right)\otimes M_{32}+M_{13}\otimes\sin \left(\frac{\varrho}{2}P_0 \right)
-\sin \left(\frac{\varrho}{2}P_0 \right)\otimes M_{13}\,,\\
\Delta_\mathcal{F}M_{10} =&M_{10}\otimes\cos \left(\frac{\varrho}{2}P_0 \right)+\cos \left(\frac{\varrho}{2}P_0 \right)\otimes M_{10}+M_{20}\otimes\sin \left(\frac{\varrho}{2}P_0 \right)
-\sin \left(\frac{\varrho}{2}P_0 \right)\otimes M_{20}\\&+ P_1\otimes \frac{\varrho}{2}M_{12}\cos \left(\frac{\varrho}{2}P_0 \right)
-\frac{\varrho}{2}M_{12}\cos \left(\frac{\varrho}{2}P_0 \right)\otimes P_1\\&+P_2\otimes \frac{\varrho}{2}M_{12}\sin \left(\frac{\varrho}{2}P_0 \right)
+\frac{\varrho}{2}M_{12}\sin \left(\frac{\varrho}{2}P_0 \right)\otimes P_2\,,\\
\Delta_\mathcal{F}M_{20} =&M_{20}\otimes\cos \left(\frac{\varrho}{2}P_0 \right)+\cos \left(\frac{\varrho}{2}P_0 \right)\otimes M_{20}-M_{10}\otimes\sin \left(\frac{\varrho}{2}P_0 \right)
+\sin \left(\frac{\varrho}{2}P_0 \right)\otimes M_{10}\\&+P_2\otimes \frac{\varrho}{2}M_{12}\cos \left(\frac{\varrho}{2}P_0 \right)
-\frac{\varrho}{2}M_{12}\cos \left(\frac{\varrho}{2}P_0 \right)\otimes P_2\\&-P_1\otimes \frac{\varrho}{2}M_{12}\sin \left(\frac{\varrho}{2}P_0 \right)
-\frac{\varrho}{2}M_{12}\sin \left(\frac{\varrho}{2}P_0 \right)\otimes P_1\,,\\
\Delta_\mathcal{F}M_{30} =&M_{30}\otimes 1+1\otimes M_{30}+\frac{\varrho}{2}P_3\otimes M_{12}-\frac{\varrho}{2}M_{12}\otimes P_3\,,
\end{array}
\ee
whereas   the Lie algebra sector, the antipode and the counit  are left undeformed. 
 
The homogeneous space on which the deformed symmetries act is characterised by a noncommutative $\star$-product constructed with the inverse twist element $\mathcal{F}^{-1}$. Following~\cite{Ciric:2017rnf}, we can express the twist in coordinate representation

by realising $P_\mu$ as $ -i\frac{\partial}{\partial x^\mu}$, and $M_{\mu\nu}$ as $ i(x_\mu\partial_\nu-x_\nu\partial_\mu)$, to obtain
\begin{equation}
\label{twist}
    \mathcal{F}=\exp\left\{-\frac{i\varrho}{2}\left[\partial_0\otimes(x^1\partial_2-x^2\partial_1)-(x^1\partial_2-x^2\partial_1)\otimes \partial_0\right]\right\}.
\end{equation}
The $\star$-product between two functions of the coordinates is defined as
\begin{equation}
\label{twistmult}
    (f\star g)(x)\coloneqq \mu_0\circ \mathcal{F}^{-1}(f\otimes g)(x)\,,
\end{equation}
where $\mu_0$ is the ordinary product map. From the twisted $\star$-product, the star-commutators  between coordinate functions are immediately obtained,
\begin{equation}
\label{twistcommrel}
\begin{aligned}
    &[x^1,x^0]_\star=i \varrho x^2\,,\\
    &[x^2,x^0]_\star=-i \varrho x^1\,,
\end{aligned}
\end{equation}
all the others being zero. They reproduce, as expected, the commutators~\eqn{algebra}. We will derive analogous commutation relations in the context of the bicrossproduct framework of quantum groups in the next section.

\section{The bicrossproduct construction}\label{bicro}

Given two Hopf algebras $\mathcal{X}, \mathcal{A}$, a \textit{bicrossproduct Hopf algebra},  $\mathcal{X} \vartriangleright \! \blacktriangleleft \mathcal{A}$~\cite{Majid1990PhysicsFA}, 
is the tensor product $\mathcal{X} \otimes \mathcal{A}$ with Hopf algebra operations,  
endowed with two  structure maps, a covariant right action of $\mathcal{X}$ on $\mathcal{A}$ and a covariant left coaction of $\mathcal{A}$ on  $\mathcal{X}$, respectively 
\begin{subequations}
\begin{align}
\triangleleft &: \mathcal{A}\times \mathcal{X}\rightarrow \mathcal{A},\\
\beta &: \mathcal{X}\rightarrow \mathcal{A}\otimes \mathcal{X},
\end{align}
\end{subequations}
which have to be compatible with the Hopf algebra structure. We review in App.~\ref{appB} the main definitions and properties.   
 The meaning of the symbol $ \vartriangleright \! \blacktriangleleft $ is that the first algebra acts on the second from the right, while the second coacts back on the first from the left, like a generalization of a semidirect product.

 For the case under analysis, the bicross-structure of $\mathcal{C}_\varrho(P)$ is obtained by identifying the algebra $\mathcal{X}$  with a deformation of the Minkowski spacetime, while $ \mathcal{A}$ will be the algebra of functions on the Lorentz group. Dually, the bicross-structure of the deformed universal enveloping $U(\mathfrak{p})$ is obtained by identifying   $\mathcal{X}$ with $U(\mathfrak{so}(1,3))$ and $ \mathcal{A}$ with the deformed Lie algebra of translations. 

\subsection{The bicrossproduct structure of the algebra  \texorpdfstring{$\mathcal{C}_\varrho (P)$}{}}\label{sec2}

The quantum Hopf algebra $\mathcal{C}_\varrho (P)$ is obtained deforming the commutative algebra of functions on the group manifold~\cite{Lizzi:2021dud, LSV}. It is possible to obtain the noncommutative structure by quantizing the Poisson bracket of coordinate functions  associated with  the classical $r$-matrix (details are reported in App.~\ref{appB1}), which reads\footnote{This can be quickly verified by expanding the twist~\eqref{rhotwist} up to the first order in the deformation parameter.}
\begin{equation}\label{rrho}
    r=i\varrho (P_0 \wedge M_{12}).
\end{equation}
For comparison, an analogous  analysis is summarized in App.~\ref{appB2} for the $\kappa$ case $\mathcal{C}_\kappa (P)$. 
Notice that the $\varrho$-Poincar\'e  $r$-matrix~\eqn{rrho} satisfies the classical Yang-Baxter equation, differently from  the $r$-matrix of the $\kappa$ case, which is known to satisfy a modified Yang-Baxter equation.

The noncommutative algebra of functions on the group manifold is obtained in~\cite{LSV} by quantizing the Poisson brackets~\eqn{poili}. The latter entails the $r$-matrix~\eqn{rrho},  besides  the left- and right-invariant vector fields of the Poincar\'e group which are spelled out in Eqs.~\eqn{lrfields} for convenience. 
Similarly to  the $\kappa$-Minkowski case, the coproduct,  antipode and counit of the Lorentz sector are undeformed, while they are deformed for the translations parameters. This result can be derived by imposing covariance of the commutation relations~\eqn{twistcommrel} (analogously~\eqn{kappacomm} for the $\kappa$ case)  under left  coaction of the quantum group (see \cite{LSV} for details).  One finally obtains 
 the following quantum group structure: 
\begin{subequations}\label{Crho}
\begin{align}
\left[ a^\mu, a^\nu \right] &=i \varrho [{\delta^\nu}_0 (a^2 {\delta^\mu}_1-a^1 {\delta^\mu}_2) -{\delta^\mu}_0 (a^2 {\delta^\nu}_1-a^1 {\delta^\nu}_2)]\,, \label{1.1a}
\\
\left[ {\Lambda^\mu}_\nu, {\Lambda^\varrho}_\sigma \right] &=0\,, \label{1.1b}
\\
\left[ {\Lambda^\mu}_\nu, a^\rho \right] &=i \varrho \left[{\Lambda^\varrho}_0 ({\Lambda^\mu}_1 g_{2\nu} -{\Lambda^\mu}_2 g_{1\nu} )-{\delta^\rho}_0 (\Lambda_{2\nu} {\delta^\mu}_1 -\Lambda_{1\nu} {\delta^\mu}_2) \right]\,, \label{1.1c}
\\
\Delta ({\Lambda^\mu}_\nu)&= {\Lambda^\mu}_\alpha \otimes {\Lambda^\alpha}_\nu\,, \label{1.1e}
\\
S({\Lambda^\mu}_\nu) &={(\Lambda^{-1})^\mu}_\nu\,, \label{1.1i}
\\
\varepsilon ({\Lambda^\mu}_\nu)&={\delta^\mu}_\nu\,, \label{1.1g}
\\
\Delta (a^\mu)&={\Lambda^\mu}_\nu \otimes a^\nu+ a^\mu \otimes 1\,,  \label{1.1d}
\\
S(a^\mu) &=-a^\nu {(\Lambda^{-1})^\mu}_\nu\,,  \label{1.1h}
\\
\varepsilon (a^\mu)&=0\,. \label{1.1f}
\end{align}
\end{subequations}
The $\varrho$-Minkowski spacetime  in this approach is assigned a priori. It determines its quantum group of symmetries, by requiring that it be the noncommutative algebra of continuous functions on the spacetime,  which is  covariant under the left coaction of the group:  
$x^\mu\rightarrow x^{\mu '}=\Lambda^\mu{}_\nu \otimes x^\nu +a^\mu \otimes 1$. We shall see in Sec. \ref{rhomi} a dual picture, where the same spacetime emerges as an appropriate quotient of the quantum universal enveloping algebra of the Poincar\'e algebra $\mathfrak{p}$.

\begin{proposition}\label{Proprho}   The quantum Hopf algebra $\mathcal{C}_\varrho (P)$ has  a bicrossproduct structure, namely
\begin{equation}\label{bicrosstructure}
\mathcal{C}_\varrho(P) =\mathcal{T}_\varrho^* \vartriangleright \! \blacktriangleleft \mathcal{C}(SO(1,3)),
\end{equation}
where $\mathcal{C}(SO(1,3))$ is the classical undeformed (Hopf) algebra of continuous functions over the Lorentz group 
and $\mathcal{T}_\varrho^*$ is the noncommutative (Hopf) algebra of functions of $\varrho$-Minkowski spacetime, given by:
\begin{subequations}
\begin{align}
[x^\mu, x^\nu] &=i \varrho [{\delta^\nu}_0 (x^2 {\delta^\mu}_1-x^1 {\delta^\mu}_2) -{\delta^\mu}_0 (x^2 {\delta^\nu}_1-x^1 {\delta^\nu}_2)]\,,\\
\Delta (x^\mu) &=x^\mu \otimes 1+1\otimes x^\mu\,,\label{1.3b}\\
S(x^\mu) &=-x^\mu\,,\\
\varepsilon(x^\mu) &=0 \label{1.3d}\,.
\end{align}
\end{subequations}
\end{proposition}
To prove this statement, we apply the construction described in App.~\ref{appB}. 
In order to define the right action {of $\mathcal T^*_\varrho$ on $\mathcal{C}(SO(1,3))$}, we first extend the elements $x^\mu\in \mathcal{T}_\varrho^*$ to $x^\mu \otimes 1 \equiv a^\mu \in \mathcal{T}_\varrho^* \otimes \mathcal{C}(SO(1,3))$ and the elements ${\Lambda^\alpha}_\beta \in \mathcal{C}(SO(1,3))$ to $1\otimes {\Lambda^\alpha}_\beta \equiv {\Lambda^\alpha}_\beta\in \mathcal{T}_\varrho^* \otimes \mathcal{C}(SO(1,3))$; then, on  assuming  a bicrossproduct structure on the tensor product quantum group, we   compute by means of the Hopf algebra product~\eqn{1.5a} and the coproduct~\eqn{1.3b}
\begin{equation}
[x^\rho \otimes 1,1\otimes {\Lambda^\mu}_\nu]= x^\rho \otimes {\Lambda^\mu}_\nu -x^\rho \otimes {\Lambda^\mu}_\nu - 1\otimes ({\Lambda^\mu}_\nu \triangleleft x^\rho ),
\end{equation}
but $[x^\rho \otimes 1,1\otimes {\Lambda^\mu}_\nu]$ is (minus) the commutator~\eqref{1.1c}. Therefore we obtain the right action in the form  
\begin{equation}
{\Lambda^\mu}_\nu \triangleleft x^\rho =i \varrho \left[{\Lambda^\rho}_0 ({\Lambda^\mu}_1 g_{2\nu} -{\Lambda^\mu}_2 g_{1\nu} )-{\delta^\rho}_0 (\Lambda_{2\nu} {\delta^\mu}_1 -\Lambda_{1\nu} {\delta^\mu}_2)\right]. \label{1.9b}
\end{equation}
Note that, from~\eqref{1.7}, ${\Lambda^\mu}_\nu \triangleleft 1={\Lambda^\mu}_\nu$ and $1\triangleleft 1=1$, but $1 \triangleleft x^\rho=0$.

\noindent To find the left coaction {of $\mathcal{C}(SO(1,3))$ on $\mathcal{T}^*_\varrho$}, we compute $\Delta(x^\mu\otimes 1)$ from~\eqref{1.5c} and the coproduct~\eqn{1.3b}:
\begin{equation}
\Delta(x^\mu\otimes 1)=x^\mu \otimes 1 \otimes 1 \otimes 1+1\otimes {x^\mu}^{(\bar{1})}\otimes {x^\mu}^{(\bar{2})}\otimes 1\,.
\end{equation}
Thus,  comparing with~\eqref{1.1d}, we find that ${x^\mu}^{(\bar{1})}={\Lambda^\mu}_\nu$ and ${x^\mu}^{(\bar{2})}=x^\nu$, therefore:
\begin{equation}
\beta(x^\mu) ={\Lambda^\mu}_\nu \otimes x^\nu. \label{1.9a}\\
\end{equation}
Note that $\beta(1)=1\otimes 1$.

Once the right action and left coaction are obtained,  we have  to verify that  the compatibility conditions~\eqref{1.6a}-\eqref{1.6d} hold. This is proven in App.~\ref{appproofcrho} in order not to burden the text.

It remains to check that the quantum group relations~\eqn{Crho} are reproduced by means of the bicrossproduct structure  given by 
Eqs.~\eqref{1.5a}-\eqref{1.5e}.

\noindent From~\eqref{1.5a} we have
\be
\begin{array}{r@{}l}
(1\otimes {\Lambda^\mu}_\nu)\cdot (1\otimes {\Lambda^\alpha}_\beta)- (1\otimes {\Lambda^\alpha}_\beta)\cdot (1\otimes {\Lambda^\mu}_\nu)&=1\otimes ({\Lambda^\mu}_\nu {\Lambda^\alpha}_\beta-{\Lambda^\alpha}_\beta {\Lambda^\mu}_\nu),\\
 (x^\mu \otimes 1) \cdot (x^\nu \otimes 1)-(x^\nu \otimes 1) \cdot (x^\mu \otimes 1) &=(x^\mu x^\nu \otimes 1+x^\mu \otimes (1\triangleleft x^\nu))- (\mu\leftrightarrow \nu)\\ 
&=x^\mu x^\nu \otimes 1 -x^\nu x^\mu \otimes 1,\\
(x^\alpha\otimes 1) \cdot (1\otimes {\Lambda^\mu}_\nu) &=x^\alpha \otimes {\Lambda^\mu}_\nu,\\
(1\otimes {\Lambda^\mu}_\nu) \cdot (x^\alpha\otimes 1) &=x^\alpha \otimes {\Lambda^\mu}_\nu +1\otimes ({\Lambda^\mu}_\nu \triangleleft x^\alpha);
\end{array}
\ee
from which  one finds the commutation rules~\eqn{1.1a}-\eqn{1.1c}.

\noindent From~\eqref{1.5c}  one finds
\be
\begin{array}{r@{}l}
\Delta (x^\mu \otimes 1) &=x^\mu \otimes 1\otimes 1\otimes 1+1\otimes {\Lambda^\mu}_\lambda \otimes x^\lambda \otimes 1,\\
\Delta (1\otimes {\Lambda^\alpha}_\beta ) &=1\otimes {\Lambda^\alpha}_\gamma \otimes 1 \otimes {\Lambda^\gamma}_\beta,\\
\Delta (x^\alpha \otimes {\Lambda^\mu}_\nu) &=x^\alpha\otimes {\Lambda^\mu}_\beta \otimes 1\otimes {\Lambda^\beta}_\nu +1\otimes {\Lambda^\alpha}_\beta {\Lambda^\mu}_\gamma \otimes x^\beta \otimes {\Lambda^\gamma}_\nu,
\end{array}
\ee
and Eqs.~\eqref{1.1e},~\eqref{1.1d} are recovered.
\noindent From~\eqref{1.5d}:
\begin{equation}
\varepsilon (x^\mu \otimes 1)=\varepsilon (x^\mu) \varepsilon (1),
\end{equation}
but from~\eqref{1.3d} $\varepsilon (x^\mu)=0$, so that we obtain~\eqref{1.1f} by identifying the LHS with $\varepsilon (a)$. Moreover
\begin{equation}
\varepsilon (1\otimes {\Lambda^\alpha}_\beta)=\varepsilon(1)\varepsilon({\Lambda^\alpha}_\beta),
\end{equation}
but $\varepsilon({\Lambda^\alpha}_\beta)={\delta^\alpha}_\beta $ and we find~\eqref{1.1g}. Finally 
\begin{equation}
    \varepsilon(x^\mu \otimes {\Lambda^\alpha}_\beta)=\varepsilon(x^\mu) \varepsilon( {\Lambda^\alpha}_\beta),
\end{equation}
which is  the expected result  from the homomorphism property of the counit.

\noindent The last condition to analyse is~\eqref{1.5e}. It yields
\begin{subequations}
\begin{align}
S(x^\mu \otimes 1) &=(1 \otimes S({\Lambda^\mu}_\nu) \cdot (S(x^\mu) \otimes 1)=-x^\nu \otimes {(\Lambda^{-1})^\mu}_\nu,\\
S(1\otimes {\Lambda^\mu}_\nu) &=(1\otimes {\Lambda^\mu}_\nu) \cdot (1\otimes 1)=1\otimes {(\Lambda^{-1})^\mu}_\nu,\\
S(x^\alpha\otimes {\Lambda^\mu}_\nu) &=S(x^\beta)\otimes S({\Lambda^\alpha}_\beta{\Lambda^\mu}_\nu)=-x^\beta \otimes {(\Lambda^{-1})^\alpha}_\beta{(\Lambda^{-1})^\mu}_\nu,
\end{align}
\end{subequations}
leading to~\eqref{1.1i} and~\eqref{1.1h}.
This completes the  proof of  the bicrossproduct structure~\eqref{bicrosstructure}.

\subsection{The bicrossproduct structure of the algebra $\mathcal{U}_\varrho (\mathfrak{p})$}\label{sec3}
In this section we  build a noncommutative Hopf algebra  which is dual to $\mathcal{C}_\varrho(P)$ and possesses a bicrossproduct structure. In order to distinguish it from $U_\varrho(\mathfrak{p})$, the one obtained from the twist in Sec.~\ref{twistedUrho},
 we will indicate  it with  a calligraphic notation,
 $\mathcal{U}_\varrho(\mathfrak{p})$.  To this, we will perform an invertible nonlinear change of generators relating the two algebras, which results to be  a Hopf algebra homomorphism as shown in the next section. 
 
 The algebra $\mathcal{U}_\varrho (\mathfrak{p})$ is the analogue, for $\varrho$-deformation, of the famous quantum Hopf algebra $\mathcal{U}_\kappa (\mathfrak{p})$, introduced by Majid and Ruegg in~\cite{Majid:1994cy}, that is,  the bicrossproduct dual of the $\kappa$-Poincar\'e $\mathcal{C}_\kappa(P)$ described in App.~\ref{appB2}. In  Sec.~\ref{hopfiso} we show that their construction applies to our case and reproduces our results.  

Primarily, let us  show that  the twist-related Hopf algebra  $U_\varrho (\mathfrak{p})$ defined by~\eqref{twistedrhoqg} does not have  a bicrossproduct structure, namely it is not of the form
$
U_\varrho(\mathfrak{p})=U(\mathfrak{so}(1,3)) \vartriangleright \! \blacktriangleleft T_\varrho,
$
where $U(\mathfrak{so}(1,3))$ is the undeformed universal enveloping algebra of the Lorentz sector, and $T_\varrho$ is the $\varrho$-deformed (Hopf) algebra of translations. To do this, let us denote  the generators of the Lorentz and translations sector of the tensor product above  respectively   with $m_{\mu\nu}$ and $p_\mu$ and let us   extend them to the whole tensor product algebra, by posing  
\be \label{tens}
P_\lambda= 1\otimes p_\lambda  \in U(\mathfrak{so}(1,3)) \otimes T_\varrho, \;\;\;\;\; M_{\mu\nu}= m_{\mu \nu}\otimes 1  \in U(\mathfrak{so}(1,3)) \otimes T_\varrho.
\ee
Given the coproducts for translation generators in~\eqref{twistedrhoqg} 
we deduce the relevant Hopf algebraic structures of $T_\varrho$, naturally identified as a subalgebra of $U_\varrho(\mathfrak{p})$, so to have
\begin{subequations}
\begin{align}
[p_\mu ,p_\nu] =&0,\\
\Delta_\mathcal{F}p_0 =&p_0\otimes 1+1\otimes p_0,\\
\Delta_\mathcal{F}p_1 =&p_1\otimes \cos \left(\frac{\varrho}{2}p_0 \right)+\cos \left(\frac{\varrho}{2}p_0 \right)\otimes p_1+p_2\otimes \sin \left(\frac{\varrho}{2}p_0 \right)-\sin \left(\frac{\varrho}{2}p_0 \right)\otimes p_2,\\
\Delta_\mathcal{F}p_2 =&p_2\otimes \cos \left(\frac{\varrho}{2}p_0 \right)+\cos \left(\frac{\varrho}{2}p_0 \right)\otimes p_2-p_1\otimes \sin \left(\frac{\varrho}{2}p_0 \right)+\sin \left(\frac{\varrho}{2}p_0 \right)\otimes p_1,\\
\Delta_\mathcal{F}p_3 =&p_3\otimes 1+1\otimes p_3,\\
S(p_\mu) =& -p_\mu,\\
\varepsilon(p_\mu)= &0\,,
\end{align}
\end{subequations}
while the  Hopf algebra $U(\mathfrak{so}(1,3))$ remains undeformed. 

 We want to show that the factorization $U_\varrho(\mathfrak{p})= U(\mathfrak{so}(1,3))\otimes T_\varrho$ generated by the elements~\eqn{tens} is not compatible with the   bicrossproduct structure. Indeed, while it is possible to read the right action of $ U(\mathfrak{so}(1,3))$ on $T_\varrho$,  the left coaction of  $T_\varrho$ is ill-defined. For the right action we apply    Eq.~\eqn{1.7}, 
 \be \label{MP}
 \left[M_{\mu\nu},P_\lambda\right]= m_{\mu\nu}\otimes p_\lambda -\bigl(m_{\mu\nu}\otimes(p_\lambda \triangleleft 1)+ 1\otimes (p_\lambda\triangleleft m_{\mu\nu})\bigr)\,,
 \ee
where we used the undeformed coproduct $\Delta m_{\mu\nu}= m_{\mu\nu}\otimes 1 + 1\otimes m_{\mu\nu}$. On the other hand, the Lie algebra sector of $U_\varrho(\mathfrak{p})$ is undeformed, so that 
\be\label{undefMP}
\left[M_{\mu\nu},P_\lambda\right]=i(g_{\nu\lambda} P_\mu -g_{\mu \lambda} P_\nu).
\ee
Comparing the two, we finally obtain 
\begin{equation}
p_\lambda\triangleleft m_{\mu \nu}=-i(g_{\nu\lambda} p_\mu -g_{\mu \lambda} p_\nu).\label{2.4}
\end{equation}
For the left coaction, however, we encounter a problem. Let us rewrite the last of Eqs.~\eqn{twistedrhoqg}
by expanding  the tensor product, as in~\eqn{tens}. We have:
\begin{equation}
\Delta(m_{30} \otimes 1)=m_{30}\otimes 1\otimes 1 \otimes 1+1\otimes 1 \otimes m_{30}\otimes 1-\frac{\varrho}{2}1\otimes p_3\otimes m_{12}\otimes 1+\frac{\varrho}{2}m_{12}\otimes 1 \otimes 1 \otimes p_3.\label{2.5}
\end{equation}
Comparing the latter with Eq.~\eqref{1.5c}, we obtain
\begin{equation}
\Delta(m_{30} \otimes 1) =(m_{30(1)}\otimes {m_{30(2)}}^{(\bar{1})})\otimes ({m_{30(2)}}^{(\bar{2})}\otimes 1),\label{2.6}
\end{equation}
where we used the notation  $\beta (x)=x^{(\bar{1})}\otimes x^{(\bar{2})}$, {with $x, x^{(\bar{2})} \in {U}(\mathfrak{so}(1,3))$}, $x^{(\bar{1})} \in {T}_\varrho, $  and  $\beta$ the sought left coaction.
It is easy to see that there is no  left coaction of ${T}_\varrho$ on $U(\mathfrak{so}(1,3)$, since the last term in Eq.~\eqref{2.5} has $p_3$ as fourth component of the tensor product, while all terms in~\eqref{2.6} have the identity as fourth component.

This means that, although $\mathcal{C}_\varrho(P)$ and $U_\varrho(\mathfrak{p})$  are dual in the Hopf-algebraic sense, they are not bicrossproduct-dual. 

In what follows we show that it is  possible to obtain, through a nonlinear change of generators of $U_\varrho(\mathfrak{p})$, a new quantum group that is not only Hopf-dual to $\mathcal{C}_\varrho(P)$, but also bicrossproduct-dual, having a bicossproduct structure itself. As already mentioned, this is analogous to the $\kappa$-Poincaré case, where the quantum group $\mathcal{C}_\kappa(P)$ is Hopf-dual to both the quantum groups of the universal enveloping algebra in the standard and Majid-Ruegg bases, but bicrossproduct-dual only to that of the Majid-Ruegg basis.

To do this, let us rewrite the Poincaré generators of $U_\varrho(\mathfrak{p})$ in a non-covariant way by defining $R_i=\frac{1}{2}\epsilon_{ijk}M_{jk}$, $N_i=M_{i0}$, so that the algebra reads:
\begin{equation}\label{poincalgebra}
\begin{matrix}
    [P_\mu,P_\nu]=0, &
    [N_i,P_0]=iP_i,  & [N_i,P_j]=i\delta_{ij}P_0,  & [N_i,N_j]=-i\epsilon_{ijk}R_k,\\
    [R_i,P_0]=0, &
    [R_i,P_j]=i\epsilon_{ijk}P_k, & [R_i,N_j]=i\epsilon_{ijk}N_k, & [R_i,R_j]=i\epsilon_{ijk}R_k.
\end{matrix}
\end{equation}
The twisted  coproducts in~\eqref{twistedrhoqg} become then

\be\label{2.1j}
\begin{array}{r@{}l}
\Delta_\mathcal{F}P_0 =&P_0\otimes 1+1\otimes P_0,\\
\Delta_\mathcal{F}P_1 =&P_1\otimes \cos \left(\frac{\varrho}{2}P_0 \right)+\cos \left(\frac{\varrho}{2}P_0 \right)\otimes P_1+P_2\otimes \sin \left(\frac{\varrho}{2}P_0 \right)-\sin \left(\frac{\varrho}{2}P_0 \right)\otimes P_2,\\
\Delta_\mathcal{F}P_2 =&P_2\otimes \cos \left(\frac{\varrho}{2}P_0 \right)+\cos \left(\frac{\varrho}{2}P_0 \right)\otimes P_2-P_1\otimes \sin \left(\frac{\varrho}{2}P_0 \right)+\sin \left(\frac{\varrho}{2}P_0 \right)\otimes P_1,\\
\Delta_\mathcal{F}P_3 =&P_3\otimes 1+1\otimes P_3,\\
\Delta_\mathcal{F}R_{1} =&R_{1}\otimes\cos \left(\frac{\varrho}{2}P_0 \right)+\cos \left(\frac{\varrho}{2}P_0 \right)\otimes R_{1}+R_{2}\otimes\sin \left(\frac{\varrho}{2}P_0 \right)-\sin \left(\frac{\varrho}{2}P_0 \right)\otimes R_{2},\\
\Delta_\mathcal{F}R_{2} =&R_{2}\otimes\cos \left(\frac{\varrho}{2}P_0 \right)+\cos \left(\frac{\varrho}{2}P_0 \right)\otimes R_{2}-R_{1}\otimes\sin \left(\frac{\varrho}{2}P_0 \right)+\sin \left(\frac{\varrho}{2}P_0 \right)\otimes R_{1},\\
\Delta_\mathcal{F}R_{3} =&R_{3}\otimes 1+1\otimes R_{3},\\
\Delta_\mathcal{F}N_{1} =&N_{1}\otimes\cos \left(\frac{\varrho}{2}P_0 \right)+\cos \left(\frac{\varrho}{2}P_0 \right)\otimes N_{1}+N_{2}\otimes\sin \left(\frac{\varrho}{2}P_0 \right)-\sin \left(\frac{\varrho}{2}P_0 \right)\otimes N_{2}\\
&+P_1\otimes \frac{\varrho}{2}R_{3}\cos \left(\frac{\varrho}{2}P_0 \right)-\frac{\varrho}{2}R_{3}\cos \left(\frac{\varrho}{2}P_0 \right)\otimes P_1+P_2\otimes \frac{\varrho}{2}R_{3}\sin \left(\frac{\varrho}{2}P_0 \right)\\
&+\frac{\varrho}{2}R_{3}\sin \left(\frac{\varrho}{2}P_0 \right)\otimes P_2,\\
\Delta_\mathcal{F}N_{2} =&N_{2}\otimes\cos \left(\frac{\varrho}{2}P_0 \right)+\cos \left(\frac{\varrho}{2}P_0 \right)\otimes N_{2}-N_{1}\otimes\sin \left(\frac{\varrho}{2}P_0 \right)+\sin \left(\frac{\varrho}{2}P_0 \right)\otimes N_{1}\\
&+P_2\otimes \frac{\varrho}{2}R_{3}\cos \left(\frac{\varrho}{2}P_0 \right)-\frac{\varrho}{2}R_{3}\cos \left(\frac{\varrho}{2}P_0 \right)\otimes P_2-P_1\otimes \frac{\varrho}{2}R_{3}\sin \left(\frac{\varrho}{2}P_0 \right)\\
&-\frac{\varrho}{2}R_{3}\sin \left(\frac{\varrho}{2}P_0 \right)\otimes P_1,\\
\Delta_\mathcal{F}N_{3} =&N_{3}\otimes 1+1\otimes N_{3}+\frac{\varrho}{2}P_3\otimes R_{3}-\frac{\varrho}{2}R_{3}\otimes P_3.
\end{array}
\ee

Inspecting Eq.~\eqn{2.6} one infers  that, in order to have a well defined left coaction,  a new  basis is needed such that   the new coproduct  has all the translation generators   on one side of the tensor product,  while the Lorentz ones are on the other side.  
To do this, we perform the  following non-linear transformation of the generators:
\be\label{redef}
\begin{array}{r@{}l}
    \widetilde{P}_0&=P_0,\\
    \widetilde{P}_1&=P_1\cos(\frac{\varrho}{2}P_0)-P_2\sin(\frac{\varrho}{2}P_0), \\
    \widetilde{P}_2&=P_2\cos(\frac{\varrho}{2}P_0)+P_1\sin(\frac{\varrho}{2}P_0), \\
    \widetilde{P}_3&=P_3,\\
    \widetilde{R}_1&=R_1\cos(\frac{\varrho}{2}P_0)-R_2\sin(\frac{\varrho}{2}P_0), \\
    \widetilde{R}_2&=R_2\cos(\frac{\varrho}{2}P_0)+R_1\sin(\frac{\varrho}{2}P_0), \\
    \widetilde{R}_3&=R_3,\\
    \widetilde{N}_1&={N}_1\cos(\frac{\varrho}{2}P_0)-{N}_2\sin(\frac{\varrho}{2}P_0)+\frac{\varrho}{2}R_3 \widetilde{P}_1, \\
    \widetilde{N}_2&={N}_2\cos(\frac{\varrho}{2}P_0)+{N}_1\sin(\frac{\varrho}{2}P_0)+\frac{\varrho}{2}R_3 \widetilde{P}_{2},\\
    \widetilde{N}_3&={N}_3+\frac{\varrho}{2}R_3P_3,
\end{array}
\ee
which is such that the Lie algebra relations continue to be undeformed. 
Details of the computation of the commutators are reported in \cref{append}. 
Let us prove that we have the desired property for the coproducts.

Exploiting the fact that
\begin{align}
       &\Delta\cos(x)=\cos(x)\otimes\cos(x)-\sin(x)\otimes\sin(x),\\
       &\Delta\sin(x)=\cos(x)\otimes\sin(x)+\cos(x)\otimes\sin(x),
\end{align}
for $x=\varrho P_0/2$, we can write the coproducts of the new generators as 
\be\label{bicrosscoprod}
\begin{array}{r@{}l}
    &\Delta \widetilde{P}_0=\widetilde{P}_0\otimes 1+1\otimes \widetilde{P}_0,\\
    &\Delta \widetilde{P}_1=\widetilde{P}_1\otimes 1 +\cos(\varrho \widetilde{P}_0)\otimes \widetilde{P}_1-\sin(\varrho \widetilde{P}_0)\otimes \widetilde{P}_2,\\
    &\Delta \widetilde{P}_2=\widetilde{P}_2\otimes 1 +\cos(\varrho \widetilde{P}_0)\otimes \widetilde{P}_2+\sin(\varrho \widetilde{P}_0)\otimes \widetilde{P}_1,\\
    &\Delta \widetilde{P}_3=\widetilde{P}_3\otimes 1+1\otimes \widetilde{P}_3,\\
    &\Delta \widetilde{R}_1=\widetilde{R}_1\otimes 1 +\cos(\varrho \widetilde{P}_0)\otimes \widetilde{R}_1-\sin(\varrho \widetilde{P}_0)\otimes \widetilde{R}_2,\\
    &\Delta \widetilde{R}_2=\widetilde{R}_2\otimes 1 +\cos(\varrho \widetilde{P}_0)\otimes \widetilde{R}_2+\sin(\varrho \widetilde{P}_0)\otimes \widetilde{R}_1,\\ 
    &\Delta \widetilde{R}_3=\widetilde{R}_3\otimes 1+1\otimes \widetilde{R}_3,\\
    &\Delta \widetilde{N}_1=\widetilde{N}_1\otimes 1 +\cos(\varrho \widetilde{P}_0)\otimes \widetilde{N}_1-\sin(\varrho \widetilde{P}_0)\otimes \widetilde{N}_2+\varrho \widetilde{P}_1\otimes \widetilde{R}_3,\\
    &\Delta \widetilde{N}_2=\widetilde{N}_2\otimes 1 +\cos(\varrho \widetilde{P}_0)\otimes \widetilde{N}_2+\sin(\varrho \widetilde{P}_0)\otimes \widetilde{N}_1+\varrho \widetilde{P}_2\otimes \widetilde{R}_3,\\
    &\Delta \widetilde{N}_3=\widetilde{N}_3\otimes 1 +1\otimes \widetilde{N}_3+\varrho \widetilde{P}_3\otimes \widetilde{R}_3.
\end{array}
\ee
Since the counits for the old generators are all zero, it is easy to see that the same result holds for the  new generators:
\begin{equation}\varepsilon(\widetilde{P}_\mu)=\varepsilon(\widetilde{N}_i)=\varepsilon(\widetilde{R}_i)=0.
\end{equation}
For the antipodes, by means of direct calculation, one finds:
\be\begin{array}{r@{}l}
    &S(\widetilde{P}_0)=-\widetilde{P}_0, \\
    &S(\widetilde{P}_1)=-\widetilde{P}_1\cos(\varrho \widetilde{P}_0)-\widetilde{P}_2\sin(\varrho \widetilde{P}_0), \\ &S(\widetilde{P}_2)=-\widetilde{P}_2\cos(\varrho \widetilde{P}_0)+\widetilde{P}_1\sin(\varrho \widetilde{P}_0), \\
    &S(\widetilde{P}_3)=-\widetilde{P}_3, \\
    &S(\widetilde{R}_1)=-\widetilde{R}_1\cos(\varrho \widetilde{P}_0)-\widetilde{R}_2\sin(\varrho \widetilde{P}_0), \\ &S(\widetilde{R}_2)=-\widetilde{R}_2\cos(\varrho \widetilde{P}_0)+\widetilde{R}_1\sin(\varrho \widetilde{P}_0),\\
    &S(\widetilde{R}_3)=-\widetilde{R}_3, \\
    &S(\widetilde{N}_1)=-\cos(\varrho \widetilde{P}_0)\widetilde{N}_1-\sin(\varrho \widetilde{P}_0)\widetilde{N}_2+\varrho\cos(\varrho \widetilde{P}_0)\widetilde{P}_1\widetilde{R}_3+\varrho\sin(\varrho \widetilde{P}_0)\widetilde{P}_2\widetilde{R}_3,\\
    &S(\widetilde{N}_2)=-\cos(\varrho \widetilde{P}_0)\widetilde{N}_2+\sin(\varrho \widetilde{P}_0)\widetilde{N}_1+\varrho\cos(\varrho \widetilde{P}_0)\widetilde{P}_2\widetilde{R}_3-\varrho\sin(\varrho \widetilde{P}_0)\widetilde{P}_1\widetilde{R}_3,\\
    &S(\widetilde{N}_3)=-\widetilde{N}_3+\varrho \widetilde{R}_3\widetilde{P}_3.
\end{array}
\ee
Therefore we have obtained a new Hopf algebra,   which we  call $\mathcal{U}_\varrho(\mathfrak{p})$, to distinguish it from the twisted one, $U_\varrho(\mathfrak{p})$.

\begin{proposition}\label{proposition2}
The quantum Hopf algebra    $\mathcal{U}_\varrho(\mathfrak{p})$ admits the bicrossproduct decomposition 
\begin{equation}
\mathcal{U}_\varrho(\mathfrak{p}) =U(\mathfrak{so}(1,3))\vartriangleright \! \blacktriangleleft \mathcal{T}_\varrho,
\end{equation}
with $\mathcal{T}_\varrho$ the new deformed translations  sector.
\end{proposition}
The proof of  the statement proceeds in two steps. One first has to find  the right action of $U(\mathfrak{so}(1,3))$ on the translations and the left coaction of 
$\mathcal{T}_\varrho$ on the Lorentz sector. Then one has  to verify the compatibility conditions of these maps, Eqs.~\eqref{1.6a}-\eqref{1.6d},  having checked  the Hopf algebra structure given by Eqs.~\eqref{1.5a}-\eqref{1.5e}.

Let us denote the standard Poincaré rotation and boost generators of $U(\mathfrak{so}(1,3))$  by $m_k,n_k$, respectively and the generators of $\mathcal{T}_\varrho$ by $p_\mu$. We then  identify $\mathcal{X}\equiv U(\mathfrak{so}(1,3))$, $\mathcal{A}\equiv \mathcal{T}_\varrho$ 

and we  extend  the generators of the two Hopf algebras  to the whole tensor product by posing
\begin{equation}
    \widetilde{M}_{\mu\nu}=m_{\mu\nu}\otimes 1,\qquad
    \widetilde{N}_i=n_i\otimes 1, \qquad \widetilde{R}_i=m_i\otimes 1, \qquad \widetilde{P}_\mu=1 \otimes p_\mu.
\end{equation}
Repeating the same steps as in Eqs.~\eqn{MP},~\eqn{undefMP} and~\eqn{2.4}, we apply~\eqref{1.7} to find
\bea
\left[\widetilde{R}_i, \widetilde{P}_\mu\right]&= & m_i\otimes p_\mu -\left(m_i\otimes (p_\mu\triangleleft 1) + 1 \otimes (p_\mu\triangleleft m_i)\right)\,,\\
\left[\widetilde{N}_i, \widetilde{P}_\mu\right]&= & n_i\otimes p_\mu -\left(n_i\otimes (p_\mu\triangleleft 1) + 1 \otimes (p_\mu\triangleleft n_i)\right)\,,
\eea
where we used the undeformed coproducts $\Delta m_{i}= m_{i}\otimes 1 + 1\otimes m_{i}$, $\Delta n_{i}= n_{i}\otimes 1 + 1\otimes n_{i}$. On the other hand, the Lie algebra sector of $\mathcal{U}_\varrho(\mathfrak{p})$ is undeformed, namely
\bea\label{undeftilMP}
\left[\widetilde{R}_i, \widetilde{P}_\mu\right]&=&i  \, \delta_{\mu j} \;\varepsilon_{ijk} \widetilde{P}_k \label{undeftilMP1}\,,\\
\left[\widetilde{N}_i, \widetilde{P}_\mu\right]&=&i  \,\delta_{i \mu} \;\widetilde{P}_0 + i \, \delta_{\mu0}\; \widetilde{P}_i.  \label{undeftilMP2}
\eea
Comparing the two sets of equations, one  finally obtains the right action of the Lorentz generators  on $\mathcal{T}_\varrho$ 
\begin{equation}
\label{alg_action}
    p_\mu \triangleleft m_i= 
    - i  \, \delta_{\mu j}\; \varepsilon_{ijk} p_k,
    \qquad 
    p_\mu\triangleleft n_i=
    - i  \, (\delta_{\mu i}\; p_0  +  \delta_{\mu0}\; p_i).
\end{equation}

In order to find the left coaction of $\mathcal{T}_\varrho$ on $U(\mathfrak{so}(1,3))$ we expand the Lorentz coproducts in~\eqn{bicrosscoprod}
\be
\begin{array}{r@{}l}
\Delta(m_A\otimes 1)=& m_A\otimes 1\otimes 1\otimes 1+ 1\otimes \cos(\varrho  p_0)\otimes m_a\otimes 1- 1\otimes \sin (\varrho p_0)\otimes\varepsilon_{AB} m_B\otimes 1\,,\\
\Delta (m_3\otimes 1)=& m_3\otimes 1\otimes 1\otimes 1+ 1\otimes 1\otimes m_3\otimes 1\,,\\
\Delta(n_A\otimes 1)=& n_A\otimes 1\otimes 1\otimes 1+ 1\otimes \cos(\varrho  p_0)\otimes n_a\otimes 1- 1\otimes \sin (\varrho p_0)\otimes\varepsilon_{AB} n_B \otimes 1\\
&+ \varrho(1\otimes p_A \otimes m_3\otimes 1)\,,\\
\Delta (n_3\otimes 1)=& n_3\otimes 1\otimes 1\otimes 1+ 1\otimes 1\otimes n_3\otimes 1+ \varrho (1\otimes p_3\otimes m_3\otimes 1)\,,
\end{array}
\ee
where $A,B=1,2$. We thus  
compare with~\eqref{1.5c}, which now reads
\be\label{compar}
\begin{array}{r@{}l}
\Delta(m_{i} \otimes 1) &=(m_{i(1)}\otimes {m_{i(2)}}^{(\bar{1})})\otimes ({m_{i(2)}}^{(\bar{2})}\otimes 1)\,,\\
\Delta(n_{i} \otimes 1) &=(n_{i(1)}\otimes {n_{i(2)}}^{(\bar{1})})\otimes ({n_{i(2)}}^{(\bar{2})}\otimes 1)\,,
\end{array}
\ee
\begin{equation}
\label{alg_coaction}
    \begin{aligned}
    \beta(m_A)&=\cos(\varrho p_0)\otimes m_A-\varepsilon_{AB}\sin(\varrho p_0)\otimes m_B,\\
    \beta(m_3)&=1\otimes m_3,\\
    \beta(n_A)&=\cos(\varrho p_0)\otimes n_A-\varepsilon_{AB}\sin(\varrho p_0)\otimes n_B+\varrho p_i\otimes m_3,\\
    \beta(n_3)&=1\otimes n_3+\varrho p_3\otimes m_3.\\
    \end{aligned}
\end{equation}
The Hopf algebra structure of the tensor product  algebra $U(\mathfrak{so}(1,3))\otimes \mathcal{T}_\varrho$, with  the right action and left coaction found above, is assured by the validity of   Eqs.~\eqref{1.5a}-\eqref{1.5e}.
Indeed, Eqs.~\eqref{1.5a},~\eqref{1.5c} are certainly satisfied since we have employed them to obtain the right action and left coaction.
Eq.\eqref{1.5b} is trivially satisfied, as well as Eq.~\eqref{1.5d}, since the counits are all zero in both the factor Hopf algebras as well as in the bicrossproduct algebra.
The remaining conditions~\eqref{1.5e} are easily verified by direct calculations.

To complete the proof that  that $\mathcal{U}_\varrho (\mathfrak{p})$ has the proper bicrossproduct structure, it remains to  check the compatibility conditions Eqs.~\eqref{1.6a}-\eqref{1.6d}.
Eq.~\eqref{1.6a} is trivially satisfied since the counits of all the generators are zero, and the right action gives always terms proportional to generators.
Eq~\eqref{1.6b} can be verified by straightforward computations applying the coproduct to the actions~\eqref{alg_action}:
\begin{equation}
    \Delta(p_\lambda\triangleleft m_{\mu\nu})=-i \left(g_{\nu\lambda}\Delta(p_\mu)-g_{\mu\lambda}\Delta(p_\nu)\right).
\end{equation}
To verify Eq.~\eqref{1.6c} one has to apply  the coaction $\beta$ to the commutators of Lorentz generators:
\begin{equation}
    \beta([m_{\mu\nu},m_{\lambda\sigma}])=i [g_{\mu\sigma} \beta(m_{\nu\lambda})-g_{\nu\sigma}\beta(m_{\mu\lambda})+g_{\nu\lambda}\beta(m_{\mu\sigma})-g_{\mu\lambda}\beta(m_{\nu\sigma})].
\end{equation}
Then, a  straightforward evaluation of  all possible Lorentz commutators shows   that~\eqref{1.6c} holds.
Eq.\eqref{1.6d} is trivially satisfied since the starting coproducts of $U(so(1,3))$ are cocommutative and the terms $a\triangleleft x_{(2)}$ in the LHS, being proportional to $p$'s, always commute with~${x_{(1)}}^{(\bar{1})}$.

This concludes the proof of Prop. \ref{proposition2}. 

\subsection{The algebra isomorphism between $\mathcal{U}_\varrho(\mathfrak{p})$ and ${U}_\varrho(\mathfrak{p})$}\label{hopfiso}

In this subsection we will follow the steps described in~\cite{Majid:1994cy} 
to show that the map between $\mathcal{U}_\varrho(\mathfrak{p})$ and ${U}_\varrho(\mathfrak{p})$ is an algebra homomorphism (indeed an isomorphism, it being invertible). 
Moreover, we shall obtain the $\varrho$-Minkowski spacetime as $\mathcal{T}_\varrho^*$,  dual  to the enveloping algebra of translations,  $\mathcal{T}_\varrho$, which is also an algebra, on which $\mathcal{T}$ and the whole $\mathcal{U}_\varrho(\mathfrak{p})$ act covariantly.

At first we consider the classical Poincar\'e algebra, which has the structure of  a semidirect sum,
$\mathfrak{g}=\mathfrak{h}\oplus_S \mathfrak{f}$,
 with $\mathfrak{h}$ the algebra of translations and $\mathfrak{f}$ the Lorentz algebra (notice that here we follow the  notation used by mathematicians with  the abelian algebra, which is acted upon by the Lorentz algebra, on the left). The mathematical structure of the semidirect sum is that of a `split extension' namely a short exact sequence  
\begin{equation}
    \mathfrak{h} \hookrightarrow \mathfrak{g} \rightarrow \mathfrak{f},
\end{equation}
with the  two maps, an inclusion $i: \mathfrak{h}\rightarrow \mathfrak{g}$ and a projection $\pi: \mathfrak{g}\rightarrow \mathfrak{f}$, such that  $\mathfrak{h}$ is an invariant subalgebra of $\mathfrak{g}$ and $\mathfrak{f}$ is   the quotient of $\mathfrak{g}$ by $\mathfrak{h}$ . The same holds for Lie groups. 
A Lie group $G$  is  a semidirect product of $H$ and $F$, namely $G=H\rtimes F$, iff
\begin{equation}
    H\hookrightarrow G\rightarrow F,
\end{equation}
is a split short exact sequence.

In~\cite{Majid1990PhysicsFA,matched,singer} a generalization is provided for quantum groups. A bicrossproduct Hopf algebra is there constructed  as a split  extension of Hopf algerbas. The approach is applied in~\cite{Majid:1994cy} to obtain the  $\kappa$-Poincar\'e quantum group $\mathcal{U}_\kappa(\mathfrak{p})$ (with the the so-called Majid-Ruegg basis) and a homomorphism is established with the $\kappa$-Poincar\'e ${U}_\kappa(\mathfrak{p})$ (with the so called standard basis). In the following we adapt  the procedure to the $\varrho$-Poincar\'e case. 

The starting observation is that $ \mathcal{U}_\varrho(\mathfrak{p})$ contains $\mathcal{T}_\varrho$ as a sub-Hopf Algebra, which can be included in $\mathcal{U}_\varrho(\mathfrak{p})$ via an inclusion map $i$. In a complementary way, $\mathcal{U}_\varrho(\mathfrak{p})$ projects onto the classical $U(\mathfrak{so}(1,3))$ with a Hopf Algebra map $\pi$:
\begin{equation}\mathcal{T}_\varrho\xhookrightarrow{i}\mathcal{U}_\varrho(\mathfrak{p})\xrightarrow{\pi} U(\mathfrak{so}(1,3)).
\end{equation}
From an operative point of view, $i$ is a standard immersion map while $\pi$ is such that 
\begin{equation}
    \pi(\widetilde{N}_i)=n_i, \qquad \pi(\widetilde{R}_i)=m_i, \qquad \pi(\widetilde{P}_\mu)=0, 
\end{equation}
where $\widetilde{N}_i,\widetilde{R}_i,\widetilde{P}_\mu\in \mathcal{U}_\varrho(\mathfrak{p})$ and $n_i, m_i\in U(\mathfrak{so}(1,3))$, as   in Sec.~\ref{sec3}.
Namely, $\pi$ projects the generators of the deformed enveloping algebra into their classical Lorentz counterparts. 

To implement the bicrossproduct construction it is also necessary to define a Hopf Algebra homomorphism $j$ and a linear map $p$ that play  an inverse role with respect to $i,\pi$. Namely
\begin{equation}\mathcal{T}_\varrho\xleftarrow{p}\mathcal{U}_\varrho(\mathfrak{p})\xhookleftarrow{j} U(\mathfrak{so}(1,3)),
\end{equation}
with
\begin{equation}
    \qquad \pi\circ j= id, \hspace{0.5cm} p\circ i=id.
\end{equation}
The linear map $p$ has to  to be a co-algebra homomorphism, meaning that
\begin{equation}
    (p\otimes p)\circ \Delta= \Delta\circ p, \qquad \varepsilon\circ p= p.
\end{equation}
Furthermore, if $j,p$ satisfy the properties
\begin{align}
    &(id\otimes j)\circ \Delta = (\pi\otimes id)\circ \Delta \circ j,\label{prop1}\\
    &p(u)t=p(u\, i(t)), \qquad u\in \mathcal{U}_\varrho(\mathfrak{p}), \qquad t\in\mathcal{T}_\varrho,\label{prop2}
\end{align}
then 
$\mathcal{U}_\varrho(\mathfrak{p})$ is a bicrossproduct extension of $U(\mathfrak{so}(1,3))$ by $\mathcal{T}_\varrho$\cite{Majid1990PhysicsFA,matched,singer}. Let us analyse the construction in some details.
By defining
\be\label{appareepsilon}
\begin{array}{r@{}l}
    \widetilde {R}_A &\coloneqq  j(m_A)=R_A\cos(\frac{\varrho}{2}P_0)-\epsilon_{AB}R_B\sin(\frac{\varrho}{2}P_0)\,, \\
    \widetilde R_3 &\coloneqq  j(m_3)=R_3\,,\\
    \widetilde N_A &\coloneqq  j(n_A)=N_A\cos(\frac{\varrho}{2}P_0)-\epsilon_{AB}N_B\sin(\frac{\varrho}{2}P_0)
    +\frac{\varrho}{2}R_3\left(P_A\cos(\frac{\varrho}{2}P_0)-\epsilon_{AB}P_B\sin(\frac{\varrho}{2}P_0)\right)\,,\\
    \widetilde N_3 &\coloneqq  j(n_3)=N_3+\frac{\varrho}{2}R_3P_3\,,
\end{array}
\ee
where, as in the previous section, $\epsilon_{AB}\,, A,B=1,2$ is the Levi-Civita pseudotensor in $2$ dimensions, one can verify that Eq.~\eqref{prop1} is satisfied. Moreover, Eq.~\eqref{prop2} holds trivially. Let us recall that in Eqs.~\eqn{appareepsilon} the lowercase generators are the ones of the classical $U(\mathfrak{so}(1,3))$,  with undeformed commutators and coproducts,  the uppercase generators with tilde, on the LHS,  are the generators of the quantum Hopf algebra $\mathcal{U}_\varrho(\mathfrak{p})$ and the uppercase generators on the RHS are related to ${U}_\varrho(\mathfrak{p})$.
In other words, the map $j$ defined by Eqs.~\eqn{appareepsilon} is nothing but the nonlinear change of basis that we have performed in the previous section from ${U}_\varrho(\mathfrak{p})$ to $\mathcal{U}_\varrho(\mathfrak{p})$. 

As for the right action and left coaction, they are given in this approach by the following definitions  
\bea
    t\triangleleft h &=& j(Sh_{(1)})tj(h_{(2)}), \qquad t\in\mathcal{T}_\varrho,\,\ h\in U(\mathfrak{so}(1,3)), \label{ract}\\
    \beta(\pi(u))&=& p(u_{(1)})Sp(u_{(3)})\otimes\pi(u_{(2)}), \qquad u\in \mathcal{U}_\varrho(\mathfrak{p}),\label{lcoact}
\eea
where the notation $a_{(3)}$ comes from the square of the coproduct $\Delta^2a=a_{(1)}\otimes a_{(2)} \otimes a_{(3)}$.

Equipped with these maps, the set $(\mathcal{T}_\varrho,U(\mathfrak{so}(1,3)),\triangleleft,\beta)$ completely determines the Hopf algebra structure of $\mathcal{U}_\varrho(\mathfrak{p})$. In fact, the commutators and coproducts are determined by the cross relations
\begin{equation} \label{hopstruc}
    \begin{aligned}
    i(t)j(h)&=j(h_{(1)})i(t\triangleleft h_{(2)}),\\
    \Delta(i(t))&=i(t_{(1)})\otimes i(t_{(2)}),\\ 
    \Delta (j(h))&=j(h_{(1)})(i\otimes j)\circ \beta(h_{(2)}),
    \end{aligned}
\end{equation}
where $h\in U(\mathfrak{so}(1,3))$, $t\in \mathcal{T}_\varrho$. By direct checking, it is easily verified that Eqs.~\eqn{ract},~\eqn{lcoact}, and~\eqn{hopstruc} reproduce the algebra and coalgebra sectors of $\mathcal{U}_\varrho(\mathfrak{p})$ presented in previous section. 

Therefore we can conclude that
\begin{itemize}
    \item 
The quantum Hopf algebra $\mathcal{U}_\varrho(\mathfrak{p})$ derived in the previous section by a non-linear change of basis for the twisted Hopf algebra ${U}_\varrho(\mathfrak{p})$ is a bicrossproduct algebra complying with  the definitions in \cite{Majid1990PhysicsFA,matched,singer};
\item It is the same Hopf algebra which one would obtain by rigorous mathematical construction via an  Hopf algebra extension of the undeformed $U(\mathfrak{so}(1,3))$ through the deformed Hopf algebra of translations $\mathcal{T}_\varrho$;
\item Because of that, and by virtue of Eqs.\eqn{appareepsilon}, 
the bicrossproduct algebra $\mathcal{U}_\varrho(\mathfrak{p})$ and the twisted Hopf algebra ${U}_\varrho(\mathfrak{p})$ are homomorphic as Hopf algebras (indeed isomorphic, being the non-linear map invertible).
\end{itemize}
\subsubsection {The \texorpdfstring{$\varrho$}{}-Minkowski spacetime re-derived}\label{rhomi}
Further building upon the results from~\cite{Majid:1994cy}, we construct the noncommutative spacetime upon which the Hopf algebra $\mathcal{U}_\varrho(\mathfrak{p})$ acts in a covariant way. 

In order to appreciate the different approaches, let us recall that, within the twist  framework, the covariant spacetime and its symmetries are defined at once by twisting both structures, see Sect. \ref{twistedUrho}. On the other hand, in terms of the quantum group  of functions over the group manifold, $\mathcal{C}_\varrho(P)$, the quantum spacetime is the primary object and the quantum symmetries are defined by requiring covariance of the former (or, viceversa, one can start from $\mathcal{C}_\varrho(P)$ and check the covariance of the quantum space), see Sect.~\ref{sec2}. 

In this section we follow a dual approach to the latter, which exploits the  bicrossproduct structure of $\mathcal{U}_\varrho(\mathfrak{p})$ and its duality with $\mathcal{C}_\varrho(P)$. We will show that the algebra of noncommutative coordinates can be obtained as   the dual of the algebra of translations $\mathcal{T}_\varrho$, which is precisely   $\mathcal{T}_\varrho^*$, the Hopf subalgebra of $\mathcal{C}_\varrho(P)$ that we have already encountered in the bicrossproduct factorization of $\mathcal{C}_\varrho(P)$, see Prop.~\ref{Proprho}. 

To this, let us apply the  definitions of Hopf algebra duality
\begin{equation}\label{hdual}
    \begin{aligned}
       &\langle t,xy \rangle= \langle t_{(1)},x\rangle \langle t_{(2)},y \rangle,\\
       & \langle ts,x\rangle =\langle t,x_{(1)}\rangle \langle s,x_{(2)}\rangle, \qquad \forall t,s\in\mathcal{T}_\varrho, \,\ \forall x,y\in \mathcal{T}_\varrho^*.
    \end{aligned}
\end{equation}
Upon introducing the dual pairing between $\mathcal{T}_\varrho$ and $\mathcal{T}_\varrho^*$, with   $t_\mu= i \widetilde{P}_\mu$,
\begin{equation}
    \langle \widetilde{P}_\mu, x^\nu \rangle = -i\delta^\nu_\mu,
\end{equation}
and using the first of Eqs.~\eqn{hdual}, we can read off the commutators between coordinates  
\begin{equation} \label{rhost}
    [x^1,x^0]=i\varrho x^2, \qquad [x^2,x^0]=-i\varrho x^1, \qquad [x^3,x^\mu]=0.
\end{equation}
After pplying the second of Eqs.~\eqn{hdual}, the coproducts turn out to be
\begin{equation}
    \Delta x^\mu = x^\mu \otimes 1 + 1\otimes x^\mu .
\end{equation}
The $\varrho$-Minkowski spacetime is thus immediately recovered. Furthermore, 
within this framework it is possible to define a natural action of the $\varrho$-Poincaré generators on the coordinates $x^\mu$, still exploiting the Hopf algebra duality relations. 
The canonical action of $\mathcal{T}_\varrho$ on $\mathcal{T}_\varrho^*$ is given by
\begin{equation}\label{transac}
    t\triangleright x = \langle t,  x_{(1)} \rangle x_{(2)}\,,
\end{equation}
which in our case simply reads 
\begin{equation}
    \widetilde{P}_\mu\triangleright x^\nu=-i \delta^{\mu}_{\nu}.
\end{equation}
The action of $U(\mathfrak{so}(1,3))$ on $\mathcal{T}_\varrho^*$ can be obtained by dualizing its action on~$\mathcal{T}_\varrho$, namely:
\begin{equation}
    \langle t, h\triangleright x\rangle = \langle t\triangleleft h,x\rangle ,
\end{equation}
with  $t\in\mathcal{T}_\varrho$, $x\in \mathcal{T}_\varrho^*$ and $h\in U(\mathfrak{so}(1,3))$. Upon using Eq.~\eqref{2.4}, such action can be read off immediately from the definition above: 

\begin{equation}
    \label{lorentzact}
    \widetilde{N}_i\triangleright x^0=i  x^i, \qquad \tilde{N_i}\triangleright x^j= -i\delta^{j}_{i}  x^0, \qquad \widetilde{R}_i\triangleright x^0 = 0, \qquad \widetilde{R}_i\triangleright x^j= \epsilon_{ijk} x^k . 
\end{equation}
With these building blocks, the action on the product of two algebra elements, say, $a$ and $b$, is given by the canonical Hopf Algebra action
\begin{equation}
    h\triangleright ab= (h_{(1)}\triangleright a)(h_{(2)}\triangleright b).
\end{equation}
for any $h\in \mathcal{U}_\varrho(\mathfrak{p})$, as required for an action to be covariant.

As an example, we prove the covariance of the commutator $[x^1,x^0]=i\varrho x^2$ under the boost $\widetilde{N}_2$. 
Using~\eqref{lorentzact} and recalling the coproducts~\eqref{bicrosscoprod}, it is easy to see that
\begin{equation}
   \widetilde{N}_2\triangleright (x^1x^0)=0, \qquad \widetilde{N}_2\triangleright (x^0x^1)=-\varrho x^0, \qquad \widetilde{N}_2 \triangleright (i\varrho x^2) = \varrho x^0,
\end{equation}
so that  the covariance condition
\begin{equation}
    \widetilde{N}_2\triangleright [x^1,x^0]=i\varrho \widetilde{N}_2\triangleright x^2 
\end{equation}
is verified. 
Covariance of the commutation relations under the action of the other Lorentz generators can be verified in a similar fashion.

For the translation generators the covariance of the commutation relations
\be\label{covtra}
\widetilde{P}_\sigma \triangleright [x^\mu,x^\nu]=i\,\varrho \left(\delta^\mu_j \delta^\nu_0- \delta^\mu_0\delta^\nu_j\right) {\epsilon^{3j}}_{k}\widetilde{P}_\sigma \triangleright x^k,
\ee
can be verified by computing  the left-hand side with the coproduct~\eqref{bicrosscoprod} and subsequently using the  action of translations on the dual,~\eqn{transac}. 
As an example, we show this for
$t_2= i \widetilde{P}_2$. 

We have
\be
t_2\triangleright (x^\mu x^\nu)= i (\widetilde{P}_{2(1)}\triangleright x^\mu)
(\widetilde{P}_{2(2)}\triangleright x^\nu),
\ee
with $\widetilde{P}_{2(1)}\otimes \widetilde{P}_{2(2)}=\widetilde{P}_2\otimes 1+ \cos(\varrho \widetilde{P}_0) \otimes \widetilde{P}_2+ \sin(\varrho P_0) \otimes \widetilde{P}_1$. We thus apply Eq.~\eqn{transac} to each term of the product, to find
\be
t_2\triangleright (x^\mu x^\nu)= -\left(\delta^\mu_2 x^\nu+  \delta^\nu_2 x^\mu+\varrho\delta_0^\mu\delta^\nu_1\right),
\ee
namely 
\be
t_2\triangleright (x^\mu x^\nu-x^\nu x^\mu)= - \varrho\left(\delta_0^\mu\delta^\nu_1-\delta_0^\nu\delta^\mu_1\right),
\ee
which is the same result that we get by computing  the right hand side of \eqn{covtra}
\be
RHS=\,\varrho \left(\delta^\mu_j \delta^\nu_0- \delta^\mu_0\delta^\nu_j\right) {\epsilon^{3j}}_{k} \delta^k_2= \varrho \left(\delta_1^\mu\delta^\nu_0-\delta_1^\nu\delta^\mu_0\right).
\ee
The proof can be repeated in a similar way for the other generators of translations. Therefore  \eqn{covtra} holds true.

\section{Plane waves and $\star$-product in \texorpdfstring{$\varrho$}{}-Minkowski}
\label{planew}
In the previous sections we have seen that the $\varrho$-Minkowski spacetime can be interpreted in two different ways. From the perspective of the twist construction, the commutation rules of spacetime coordinates are computed by means of the $\star$-product corresponding to  the twist that generates the algebra of symmetries discussed in section \ref{twistedUrho}. From the perspective of the bicrossproduct construction, the $\varrho$-Minkowski commutation relations are deduced by exploiting the duality between the translation generators and spacetime coordinates. 

On the other hand, starting from the commutation rules of spacetime one can represent the algebra of noncommutative functions as an algebra of operators by choosing a specific ordering for   the basis of noncommutative plane waves $\hat\phi(p)=:e^{i p_\mu \hat x^\mu}:$~\cite{Amelino-Camelia:1999jfz}.   
The operator product  of   plane waves $\hat \phi(p)$  defines a noncommutative $\star$-product for functions on spacetime, 
according to
\be \label{starpp}
(f\star g)(x) =\int \dd^4 p \, \dd^4 k\, \Tilde{f}(p)\Tilde{g}(k) \; \langle\hat\phi(p) \hat\phi(k)\rangle\,,
\ee
where the tilde indicates the standard Fourier transform.  Different ordering prescriptions will define different products. In App.~\ref{appe} a short review for the analogous case of  $\kappa$-Minkowski is given, in order to highlight the similarities. 
The notation
$
\langle\hat\phi(p) \hat\phi(k)\rangle
$ defines  the noncommuting function, 
$\phi(p\oplus k)= \exp\left[i(p\oplus k)\cdot x\right]$, which corresponds to the operator between brackets once the operator product is computed, $\oplus$ representing a deformed sum. 
In terms of noncommutative exponentials $
e^{i p_\mu  x^\mu}$ 
the $\star$-product between  functions reads, then,\footnote{We ignore here convergence issues. The product is well defined for a fairly large class of functions, which includes the Schwarzian ones.}
\be \label{starp}
(f\star g)(x) =\int \dd^4 p \, \dd^4 k\, \Tilde{f}(p)\Tilde{g}(k) \; \rm{e}^{ip_\mu x^\mu}
\star  \rm{e}^{ik_\nu x^\nu}\,.
\ee
with 
\be\label{expprod}
e^{ip\cdot x}\star e^{ik\cdot x}=e^{i(p\oplus k)\cdot x}.
\ee
Moreover, the  composition rule of plane waves {\it defines}  a  coproduct for the translation generators. This  will be derived by requiring consistency of the action of translations  on both sides of \eqn{expprod}, or, equivalently, by acting on both sides of the operator version  $\hat{\phi}(p)\hat{ \phi}(k)= \hat \phi(q)$,  with
$q=p\oplus k$. 
The  operator $\hat \phi(q)$ and the related function 
will be different for different ordering prescriptions. Therefore, the deformed sum $\oplus$ defines   different  coproducts for the translation generators, as we shall see in detail below. 

The question arises as to whether the $\star$-product and coproduct derived from plane wave multiplication 
correspond to any of the structures defined either through the  bicrossproduct  or  the twist constructions. 

Different ordering prescriptions of the plane waves result in different coproducts of the translation generators, and in different $\star$-products. 
We shall show that in fact only a specific ordering choice, the so called {\it time-to-the-right}  prescription, results in a coproduct that is compatible with translation generators associated with  the bicrossproduct construction. This ordering also defines a novel $\star$-product  which appears to be different from previously known ones \cite{DimitrijevicCiric:2018blz, Hersent:2023lqm}. Other ordering choices correspond to different coproducts, that is to non-linear redefinitions of the translation generators.    
In particular,  the so called {\it time-symmetric} prescription is associated with  the translation generators in the twisted basis, and it coherently  defines the  $\star$-product which is obtained from the twist \cite{DimitrijevicCiric:2018blz}.

In order to define plane waves and their rule of multiplication, we follow~\cite{Freidel:2007hk, Kowalski-Glikman:2013rxa} and  start with  (finite dimensional) representations of the spacetime algebra.
The commutation relations~\eqref{algebra} can be represented by means of $4\cross 4$ matrices
\begin{equation}
\label{corrrep}
\begin{aligned}
    \hat x^0=\begin{pmatrix}
        0 & i\varrho & 0 & 0 \\
        -i\varrho & 0 & 0 & 0 \\
        0 & 0 & 0 & 0 \\
        0 & 0 & 0 & 0
    \end{pmatrix} , \quad 
      \hat x^1=\begin{pmatrix}
        0 & 0 & 0 & i\varrho  \\
        0 & 0 & 0 & 0 \\
        0 & 0 & 0 & 0 \\
        0 & 0 & 0 & 0
    \end{pmatrix} ,\\
        \hat x^2=\begin{pmatrix}
        0 & 0 & 0 & 0  \\
        0 & 0 & 0 & i\varrho \\
        0 & 0 & 0 & 0 \\
        0 & 0 & 0 & 0
    \end{pmatrix} ,\quad 
            \hat x^3=\begin{pmatrix}
        0 & 0 & 0 & 0  \\
        0 & 0 & 0 & 0 \\
        0 & 0 & 0 & i\varrho \\
        0 & 0 & 0 & 0
    \end{pmatrix},
\end{aligned}
\end{equation}
which formally close the Lie algebra  of the centrally extended Euclidean group in 3 dimensions, with one rotation ($\hat x^0$) and two spatial translations ($\hat x^1,\hat x^2$), while $\hat x^3$ is a central generator.

\subsection{Time-to-the-right ordering and bicrossproduct basis}

Using the representation~\eqref{corrrep}, we can write plane waves in matrix form. Choosing the time-to-the-right ordering we define
\begin{equation}
\label{pwr}
    \hat \phi_R(p)=e^{ip_k\hat x^k}e^{ip_0\hat x^0}= \begin{pmatrix}
    \cos(\varrho p_0) & -\sin(\varrho p_0) & 0 & -\varrho p_1 \\
    \sin(\varrho p_0) & \cos(\varrho p_0) & 0 & -\varrho p_2 \\
    0 & 0 & 1 & -\varrho p_3 \\
    0 & 0 & 0 & 1
    \end{pmatrix}.
\end{equation}
Multiplying two plane waves we get
\begin{equation}
\begin{aligned}
\label{ttrsum}
    &\hat\phi_R(p)\hat\phi_R(k)=\hat \phi_R(p\oplus_R k)=\\
   & =\begin{pmatrix}
    \cos(\varrho(k_0+p_0)) & -\sin(\varrho(k_0+p_0)) & 0 & -\varrho [p_1+k_1\cos(\varrho p_0)-k_2\sin(\varrho p_0)] \\
    \sin(\varrho(k_0+p_0)) & \cos(\varrho(k_0+p_0)) & 0 & -\varrho[p_2+k_2\cos(\varrho p_0)+k_1\sin(\varrho p_0)] \\
    0 & 0 & 1 & -\varrho(p_3+k_3) \\
    0 & 0 & 0 & 1 
    \end{pmatrix}.
\end{aligned}
\end{equation}
Comparing the last column of the single plane wave representation, Eq.~\eqref{pwr}, to the last column of~\eqref{ttrsum}, we find the law of addition of momenta, 
\begin{equation}
\label{complaws}
    \begin{cases}
    (p\oplus_R k)_0= p_0+k_0,\\
    (p\oplus_R k)_1= p_1+k_1\cos(\varrho p_0)-k_2\sin(\varrho p_0),\\
    (p\oplus_R k)_2= p_2 + k_2\cos(\varrho p_0)+k_1\sin(\varrho p_0), \\
    (p\oplus_R k)_3=p_3+k_3.
    \end{cases}
\end{equation}
Therefore, the multiplication of plane waves is a new plane wave of momentum $p\oplus_R k$. Namely, it is still an eigenfunction of translations. This can in turn be related to the coproduct of translation generators $P_\mu^R$, as we now demonstrate.

In the language of noncommutative plane wave operators $\hat \phi_R=e^{ip_k\hat x^k}e^{ip_0\hat x^0}$, on which the translation generators act as $P_\mu^R \hat \phi_R(p)=p_\mu \hat \phi_R(p)$, we must have  
\be
P^R_\mu\left(\hat\phi(p)\hat \phi(k)\right)=P^R_\mu \hat\phi(p\oplus_R k) = (p\oplus_R k)_\mu \hat\phi(p\oplus_R k).
\ee
One can verify by direct calculation that,
in order for this to hold, 
the coproduct of $P_\mu^R$ should be:
 \be\label{coproductspr}
 \begin{array}{r@{}l}
    &\Delta P_0^R=P_0^R\otimes 1+1\otimes P_0^R,\\
    &\Delta P_1^R=P_1^R\otimes 1 +\cos(\varrho P_0^R)\otimes P_1^R-\sin(\varrho P_0^R)\otimes P_2^R,\\
    &\Delta P_2^R=P_2^R\otimes 1 +\cos(\varrho P_0^R)\otimes P_2^R+\sin(\varrho P_0^R)\otimes P_1^R,\\
    &\Delta P_3^R=P_3^R\otimes 1+1\otimes P_3^R\,.
    \end{array}
    \ee
 Let us check it for 
 $P_2^R$, as  the others can be easily verified analogously. On indicating with $\hat\mu$ the  standard operator product, we have, applying  \eqn{coproductspr} to the product of plane waves
\be
\begin{array}{r@{}l}
P^R_\mu\left(\hat \phi(p)\hat\phi(k)\right)&=
    \hat \mu \circ \Delta P^R_2 (\hat\phi(p)\otimes \hat\phi(k))\\
    &=\hat \mu\circ \Bigl[ P^R_2\hat\phi(p)\otimes \hat\phi(k) +\cos(\varrho P^R_0)\hat\phi(p)\otimes \hat\phi(k)+
    \sin(\varrho P^R_0)\hat\phi(p)\otimes P^R_1\hat\phi(k)\Bigr]\\
    & =\left( p_2+\cos(\varrho p_0)k_2+\sin(\varrho p_0) k_1\right)\hat \phi(p)\hat\phi(k)\,.
\end{array}
\ee
In terms of noncommutative functions $\phi(p)
= e^{i p_\nu x^\nu}$  
the translation generators  act as differential operators, $P_\mu^R \phi(p)=-i\del_\mu e^{i p_\nu x^\nu}=p_\mu \phi(p)$. \footnote{Notice that the superscript ${\;}^R$ for the generator of translations is redundant when acting on a single plane wave. We maintain it for book keeping.}
Therefore we must have 
\be
P^R_\mu \left(\phi(p)\star\phi(k)\right)=P^R_\mu \phi(p\oplus_R k)= -i \del_\mu \exp (i(p\oplus_R k)_\nu x^\nu)= (p\oplus_R k)_\mu \phi(p\oplus_R k).
\ee
Again, this holds if  
the coproduct of $P_\mu^R$  takes the form \eqn{coproductspr}.
As done before, we check it for 
 $P_2^R$. On indicating with $\mu$ the $\star$ product between functions, and applying  \eqn{coproductspr} to the product of plane waves, we have
\be
\begin{array}{r@{}l}
P^R_\mu\left(\phi(p)\star\phi(k)\right)&=
    \mu \circ \Delta P^R_2 (\phi(p)\otimes \phi(k))=\mu\circ \Bigl[ P^R_2\phi(p)\otimes \phi(k) +\cos(\varrho P^R_0) \phi(p)\otimes \phi(k)\bigr.\\
    &+\bigl.\sin(\varrho P^R_0)\phi(p)\otimes P^R_1\phi(k)\Bigr]= \left(p_2+\cos(\varrho p_0)k_2+\sin(\varrho p_0) k_1\right)\left(\phi(p)\star\phi(k)\right).
\end{array}
\ee
Comparing with the results obtained in the previous sections,  we find that the $\varrho$-Minkowski coproducts for translation operators in the bicrossproduct basis, given in~\eqref{bicrosscoprod}, take exactly the form~\eqref{coproductspr}. 
We conclude that the bicrossproduct basis is compatible with the time-to-the-right ordering for plane-waves.

We briefly mention what happens upon choosing the {\it time-to-the-left} ordering. With analogous calculations, it is possible to show that the structure of the coproducts is similar to the one obtained with the {\it time-to-the-right} ordering, namely the enveloping algebra is still a bicrossproduct, modulo an inversion of the deformation parameter (i.e. $\varrho\rightarrow-\varrho$) and a swap between the tensor product spaces, namely:
\begin{equation}
\label{copleft}
\begin{aligned}
    &\Delta P^L_1=1\otimes P^L_1 + P^L_1\otimes \cos(\varrho P^L_0)+P^L_2\otimes \sin(\varrho P_0^L),\\
    &\Delta P^L_2=1\otimes P^L_2 + P^L_2\otimes \cos(\varrho P^L_0)-P^L_1\otimes \sin(\varrho P_0^L),\\
    &\Delta P^L_{0,3}=P^L_{0,3}\otimes 1 +1\otimes P^L_{0,3} \; .
\end{aligned}
\end{equation}
An analogous result is also present when analysing the {\it time-to-the-right} and {\it time-to-the-left} plane waves in $\kappa$-Minkowski.

\subsubsection{The \texorpdfstring{$\star$}{}-product}\label{cycl}
While the $\star$-product can be defined for any ordering, in the following we concentrate on the time-to-the-right one. 
From Eqs. \eqn{starp}, \eqn{expprod}  
we find 
\bea
( f\star g) (x) &=& \int \dd^4 p\; \dd^4k \tilde{f}(p_0-k_0,p_A-R_{AB}(-\varrho p_0+\varrho k_0)k_B,p_3-k_3)\tilde{g}(k)e^{ip_\mu x^\mu}\nonumber\\
 &=& \int \dd^4 p\;  (\Tilde f\circ \Tilde g)(p)e^{i p_\mu x^\mu}\,, \label{starpr}
\eea
with $R_{AB}(\theta), \; A,B=1,2$ the rotation matrix in the 1-2 plane of argument $\theta$ and 
\be 
(\Tilde f\circ \Tilde g)(p) =\int \dd^4  k \; \Tilde f(p_0-k_0,p_A-R_{AB}(-\varrho p_0+\varrho k_0)k_B,p_3-k_3)  \;\Tilde g(k)
\ee
the deformed convolution of Fourier transforms.

Let us show by direct calculation that this product is  cyclic with respect to the standard integration measure on $\mathbb{R}^4$ namely that 
\begin{equation}
    \int \dd^4{x}\; f({x})\star g({x}) =\int \dd^4{x}\; g({x})\star f({x}).
\end{equation}
From  the expression obtained in terms of plane waves \eqn{starpr} we compute
\begin{equation}
\begin{aligned}
\int \dd^4 x \; ( f\star g) (x) = & \int \dd^4 x \; \dd^4 p\; \dd^4k \; \tilde{f}(p_0-k_0,p_A-R_{AB}(-\varrho p_0+\varrho k_0)k_B,p_3-k_3)\tilde{g}(k)e^{ip_\mu x^\mu}\\
&= \int \dd^4 p \; \dd^4k \tilde{f}(p_0-k_0,p_A-R_{AB}(-\varrho p_0+\varrho k_0)k_B,p_3-k_3)\tilde{g}(k)\delta(p_\mu)\\
&= \int \dd^4k\; \tilde{f}(-k_0,-R_{AB}(\varrho k_0),-k_3)\tilde{g}(k)\\
&= \int \dd^4k\; \tilde{f}(k)\tilde{g}(-k_0,-R_{AB}(\varrho k_0)k_B,-k_3)\\
& = \int \dd^4 p \; \dd^4k \tilde{f}(k)\tilde{g}(p_0-k_0,p_A-R_{AB}(-\varrho p_0+\varrho k_0)k_B,p_3-k_3)\delta(p_\mu)\\ 
&= \int \dd^4 x \; \dd^4 p\; \dd^4k \; \tilde{f}(k)\tilde{g}(p_0-k_0,p_A-R_{AB}(-\varrho p_0+\varrho k_0)k_B,p_3-k_3)e^{ip_\mu x^\mu}\\& =
\int \dd^4 x \; (g\star f)(x).
\end{aligned}
\end{equation}
We thus conclude that the $\star$-product derived from  the time-to-the-right ordering for plane waves is indeed cyclic. Furthermore, it is also $\varrho$-Poincaré invariant, since the measure is undeformed.

Notice however that  the 
 stronger closure condition \cite{Felder:2000nc} $\int f\star g= \int f \cdot g$ does not hold. 
 It has been already noticed that the cyclicity property is a fundamental property of the $\star$-product because it allows to formulate field theories in terms of a $\star$-gauge invariant action (namely such that the fields transform according to $\psi(x)\rightarrow g(x)\star \psi(x)$, with $g$ a unitary element of the noncommutative algebra). When the stronger closure condition is fulfilled one has the further simplifying property that quadratic terms of the action, such as the kinetic and the mass term, are undeformed, hence they produce the standard tree level propagator. This is the case of the twisted $\varrho$-Minkowski product, for which the closure property has been employed to investigate the behaviour  of scalar field theories at one loop \cite{dimi2}.  We recall that an analogous situation holds for Moyal and Wick-Voros products, which represent the same spacetime noncommutativity, both being cyclic, but only Moyal being closed. Their application to scalar field theory and a comparison between the results has been analysed  in \cite{Galluccio:2008wk}. It would be interesting to analyse  the consequences of the new product \eqn{starpr} within a similar setting. We plan to come back to applications in a further publication.\footnote{Another $\star$-product for the $\varrho$-Minkowski spacetime has been  recently found \cite{Hersent:2023lqm},  based on Weyl quantization. It appears to be  different from the ones presented here.  The latter is cyclic but not closed. It has the interesting property that $\int f^\dag \star g = \int \bar f \cdot g$ where the adjoint is defined w.r.t. a suitable integration measure.}

\subsection{Time-symmetric ordering and twist basis}
Consider the time-symmetric ordering for plane waves 
\begin{equation}
\label{pws}
\begin{aligned}
    & \hat\phi_S(p)=e^{i\frac{p_0\hat x^0}{2}}e^{ip_k\hat x^k}e^{\frac{p_0\hat x^0}{2}}=\\
    &=\begin{pmatrix}
        \cos(\varrho p_0) & -\sin(\varrho p_0) & 0 & -\varrho (p_1\cos(\varrho\frac{p_0}{2})-p_2\sin(\varrho\frac{p_0}{2})) \\
    \sin(\varrho p_0) & \cos(\varrho p_0) & 0 & -\varrho (p_2\cos(\varrho\frac{p_0}{2})+p_1\sin(\varrho\frac{p_0}{2})) \\
    0 & 0 & 1 & -\varrho p_3 \\
    0 & 0 & 0 & 1
    \end{pmatrix}.
\end{aligned}
\end{equation}
The product of symmetric ordered plane waves $\hat \phi_S(p)\hat \phi_S(k)$ is given by
\begin{equation}
  \begin{pmatrix}
  \cos(\varrho(k_0+p_0)) & -\sin(\varrho(k_0+p_0)) & 0 & -\varrho F(p_0,k_0,p_1,k_1,p_2,k_2) \\
    \sin(\varrho(k_0+p_0)) & \cos(\varrho(k_0+p_0)) & 0 & -\varrho G(p_0,k_0,p_1,k_1,p_2,k_2) \\
    0 & 0 & 1 & -\varrho(p_3+k_3) \\
    0 & 0 & 0 & 1 
    \end{pmatrix},
\end{equation}
with 
\begin{equation}
\begin{aligned}
    F(p_0,k_0,p_1,k_1,p_2,k_2)=&p_1\cos(\varrho \frac{p_0}{2})+k_1\cos(\frac{\varrho}{2}(k_0+2p_0))\\&+p_2\sin(\varrho \frac{p_0}{2})+k_2\sin(\frac{\varrho}{2}(k_0+2p_0)),\\
    G(p_0,k_0,p_1,k_1,p_2,k_2)=&p_2\cos(\varrho \frac{p_0}{2})+k_2\cos(\frac{\varrho}{2}(k_0+2p_0))\\&+p_1\sin(\varrho \frac{p_0}{2})+k_1\sin(\frac{\varrho}{2}(k_0+2p_0)).
    \end{aligned}
\end{equation}
Using trigonometric identities,  this can be  put in the form $\hat \phi_S(p\oplus_S k)$ where the deformed  composition law $\oplus_S$ is defined as follows
\begin{equation}
    \begin{cases}
        (p\oplus_S k)_0=p_0+k_0,\\
        (p\oplus_S k)_1=p_1\cos(\frac{\varrho}{2}k_0)+\cos(\frac{\varrho}{2}p_0)k_1+p_2\sin(\frac{\varrho}{2}k_0)-\sin(\frac{\varrho}{2}p_0)k_2,\\
        (p\oplus_S k)_2=p_2\cos(\frac{\varrho}{2}k_0)+\cos(\frac{\varrho}{2}p_0)k_2-p_1\sin(\frac{\varrho}{2}k_0)+\sin(\frac{\varrho}{2}p_0)k_1,\\
        (p\oplus_S k)_3=p_3+k_3.
    \end{cases}
\end{equation}
Let us notice that this is exactly the deformed sum obtained in \cite{DimitrijevicCiric:2018blz} by computing the twisted $\star$-product of plane waves.
By introducing  time-symmetric translation generators, we want to find, as in the previous subsection, the coproduct $\Delta P^S$ which is compatible with the request that $\phi(p\oplus_S k)$ be an eigenfunction of translations $P^S_\mu \phi(p\oplus_S k)=(p\oplus_S k)_\mu\phi(p\oplus_S k )$.  It turns out that this composition law can be checked to be   compatible with the twisted coproducts~\eqref{twistedrhoqg} of translation generators. Indeed, acting with the twisted coproducts on the product of two symmetric ordered plane-waves, one obtains a composition law of $p$ and $k$ that is the same as the one given by the $\oplus_S$ operation. 
Consistently, one can verify  that the $\star$-product which is obtained by convolution of  time-symmetric plane waves along the same lines as in the previous section,   matches the one that is computed in \cite{DimitrijevicCiric:2018blz} using the twist, Eq.~\eqref{twist}. We have, therefore, obtained the plane waves representation which corresponds to the twisted $\star$-product.

In the previous sections, we have shown that  the twist translation generators and bicrossproduct translation generators  are related by a nonlinear transformation. We can verify that the same relation holds between the momenta of the time-symmetric and the time-to-the-right ordered plane waves. 
To do this, starting from the momenta in the time-symmetric plane waves, $p_i$, we perform the following transformations
\begin{equation}
\begin{aligned}
\label{cob}
    \Tilde{p}_1&=p_1\cos(\varrho\frac{p_0}{2})-p_2\sin(\varrho\frac{p_0}{2}),\\ \Tilde{p}_2&=p_2\cos(\varrho\frac{p_0}{2})+p_1\sin(\varrho\frac{p_0}{2}),\\ \Tilde{p}_{0,3}&=p_{0,3} \;,
\end{aligned}
\end{equation}
which we apply to the time-symmetric plane waves \eqn{pws}. One can verify that the transformed plane waves are eigenfunctions of the translation generators
\begin{equation}
\begin{aligned}
    \widetilde{P}_1&=P^S_1\cos(\varrho\frac{P^S_0}{2})-P^S_2\sin(\varrho\frac{P^S_0}{2}),\\  \widetilde{P}_2&=P^S_2\cos(\varrho\frac{P^S_0}{2})+P^S_1\sin(\varrho\frac{P^S_0}{2}),\\
    \widetilde{P}_{0,3}&=P^S_{0,3}\; ,
\end{aligned}
\end{equation}
which are precisely the generators obtained in Sec. \ref{sec3} through the change of basis
~\eqref{redef} and one can identify $\widetilde{P}_\mu$ with  $P_\mu^R$. 

\subsubsection{Equivalence class of star products}
It is known that different $\star$-products associated with a given spacetime noncommutativity are equivalent, in the sense that an invertible map $T$ can always be found, such that
\be
T(f\star g) = Tf\star' Tg.
\ee
In the framework of phase-space quantization this redundance is well known and related to ordering ambiguities. For quantum spacetime, it has been studied for the class of translation invariant $\star$-products \cite{Galluccio:2009ss}, of which the Moyal product is the only closed representative. For rotation invariant products the issue is discussed in  \cite{Kupriyanov:2015uxa}.

Interestingly, for the $\varrho$-Minkowski case  such a map is easily found in terms of the invertible relations between 
the Poincar\'e generators related to the various orderings that it is possible to choose for the $\star$-product of plane waves. In particular, Eqs. \eqn{cob} connecting the time-symmetric ordering with the time-to-the right ordering allow to connect the twisted $\star$-product found in \cite{DimitrijevicCiric:2018blz}, which is closed,  with the bicrossproduct \eqn{starpr} which is only cyclic.
The map can be explicitly written in Fourier transform
\be
Tf := \int \dd^4 p \, \tilde f(p)  \tilde T {\rm e}^{i p_\mu x^\mu}
\ee
with $
 \tilde T {\rm e}^{i p_\mu x^\mu}:=  {\rm e}^{i \tilde p_\mu x^\mu}$ according to  \eqn{cob}.

\section{Conclusions}
The noncommutative spacetime $\varrho$-Minkowski shows some interesting features, which single it out among the variations of the original $\kappa$-Minkowski. Similarly to the latter, it has an undeformed algebra 
of Lorentz 
symmetries, $C(SO(3,1))$, as a subalgebra of the quantum Poincar\'e group $C_\varrho (P)$, as well as un undeformed Lorentz Lie algebra sector 
and a bicrossproduct construction which make it possible to read its quantum relativity group as a natural extension of the semidirect product of Lie groups to the Hopf algebraic framework. Moreover, the bicrossproduct structure allows to obtain the spacetime (i.e. the noncommutative algebra $\mathcal{T}^*_\varrho$) as a generalization of the concept of homogeneous space which holds for the Minkowski spacetime in the classical setting.  

Besides the bicrossproduct derivation, the $\varrho$-Minkowski symmetries may be obtained by a twist. In this case the algebra is  generated by an $r$-matrix which is solution of the classical Yang-Baxter equation, whereas, up to our knowledge,  the twists related to  the $\kappa$-Minkowski algebra are associated with modified Yang-Baxter equations and, as such, they require an enlargement of the Poincar\'e symmetry in order to be defined. The deformed Poincar\'e symmetries, $\mathcal{U}_\varrho\mathfrak{p})$  and $U_\varrho\mathfrak{p})$, obtained following the bicrossproduct and the twist approaches, are isomorphic quantum groups related by a nonlinear change of generators. 

Using a finite-dimensional representation of the algebra of coordinates and expanding the noncommutative functions on a basis of plane waves,  it has been possible to define different $\star$-products related to different ordering prescriptions. We have found that the {\it time-symmetric} ordering singles out the twisted star product already found in \cite{dimi2}, whereas the {\it time-to-the-right} ordering yields a novel $\star$-product, consistent with the  bicrossproduct construction. We have shown that it enjoys  the cyclicity property with respect to the standard measure in $\mathbb{R}^4$ and,  unlike the twisted product, it is   not closed. Therefore, it would be interesting to  
to study  gauge theories within the two approaches, invariance being ensured by cyclicity, and compare the results.  

There are other interesting features, such as the discretization of the spectrum of the time observable 
due to  the non-commutativity  of coordinates \cite{Lizzi:2021dud}, which could have phenomenological applications. 
Moreover, a $2+1$-dimensional version of the $\varrho$-Minkowski framework was taken as a starting point for quantum gravity phenomenology studies. In \cite{Amelino-Camelia:2016wpo,Amelino-Camelia:2017pne}, the $\varrho$-deformations of the Poincaré symmetries predict the so-called dual-lensing effect, which might open a new window on Planck-scale phenomenology. The algebra of relativistic symmetries used in these studies is however different from the ones we have discussed in this work. In particular, the Lorentz sector is deformed. An analysis   of the different implications for these models will be matter of future work. Another aspect which deserves further investigation  is the relation with the $\lambda$-Minkowski spacetime (where the role of $x^0$ and $x^3$ is exchanged), whose analysis is likely to be very similar, although the physics it describes would be different.

\section*{Acknowledgments}
We acknowledge support from the
INFN Iniziativa Specifica GeoSymQFT (F.L., P.V.) and Quagrap (G.F., G.G.). F.L.\ acknowledges financial support from the
State Agency for Research of the Spanish Ministry of Science and Innovation through
the “Unit of Excellence Maria de Maeztu 2020–2023” award to the Institute of Cosmos
Sciences (Grant No. CEX2019-000918-M) and from Grants No. PID2019–105614 GB-
C21 and No. 2017-SGR-929.  The research of G.F, G.G, F.L.\ and P.V.\  was carried out in the frame of Programme STAR Plus,
    financially supported  by UniNA and Compagnia di San Paolo. 
L.S. acknowledges financial support from the doctoral school of the University of Wrocław and the SONATA BIS grant 2021/42/E/ST2/00304
from the National Science Centre (NCN), Poland.

\appendix

\section{The twist}\label{appA}
In this appendix we give some defintions and show some properties of the twist approach.

A Drinfel'd twist $\mathcal{F}$ is an invertible map $\mathcal{F}\in U(\mathfrak{g})\otimes U(\mathfrak{g})$, with an action on the algebra of functions on the group, 
\be
\mathcal{F}: \mathcal{C}(G)\otimes \mathcal{C}(G)\rightarrow \mathcal{C}(G)\otimes \mathcal{C}(G),
\ee
which satisfies the following cocycle and normalization conditions 
\begin{align}
(\mathcal{F}\otimes 1)(\Delta \otimes id)\mathcal{F}&=(1\otimes \mathcal{F})(id \otimes \Delta)\mathcal{F},\\
(\varepsilon \otimes 1)\mathcal{F} &=(1\otimes \varepsilon)\mathcal{F}=1.
\end{align}
Twist operators that do not satisfy the cocycle condition are called \textit{non admissible twists}.

In terms of this map it is possible to introduce a noncommutative $\star$-product between functions defined on $G$ as follows
\begin{equation}
f\star g =\mu_\star (f\otimes g) \doteq \mu \circ \mathcal{F}^{-1} (f\otimes g), \hspace{0.5cm} f,g\in \mathcal{C}(G);
\end{equation}
where $\mu_\star$ is the noncommutative deformation of the classical commutative product $\mu: \mathcal{C}(G)\otimes \mathcal{C}(G) \rightarrow \mathcal{C}(G)$.
The cocycle and the normalization conditions imply that the $\star$-product is associative and the existence of the neutral element $1$: $f\star 1=1\star f=f$.

In our work we will deal with twists of a particular kind called \textit{abelian twists} (namely constructed in terms of commuting generators of the Lie algebra), and for this case it is possible to show that
\begin{equation}
\mathcal{F}^{-1}\approx 1\otimes 1+\frac{1}{2}r+\dots
\end{equation}
i.e., that the classical $r$-matrix is given by the first order in the deformation parameter expansion of the twist operator.

Connected with this property, it is possible to show that if the classical $r$-matrix does not solve a CYBE but a modified one, then the dual universal enveloping algebra does not admit an abelian admissible twist. This is the case of $\kappa$-Poincaré, and the reason we are forced to obtain its quantum algebra via other procedures, while in the $\varrho$-Poincaré case, since the $r$-matrix satisfies a CYBE, we can employ the twist procedure.

For a Lie algebra-type deformation that admits an admissible twist operator, a simple way to obtain a quantum enveloping algebra of Poincaré is achieved deforming the standard Poincaré Hopf algebra via the relation
\begin{equation}
\Delta_\mathcal{F}=\mathcal{F}\Delta \mathcal{F}^{-1},
\end{equation}
leaving all the other maps undeformed.

The $\kappa$-Minkowski algebra has been described in terms of a $\star$-product derived via many different twisting operators. These are the so called Jordanian~\cite{kulish, giaquinto, tolstoy, Borowiec} and Abelian twists~\cite{reshetikhin, bu, Govindarajan:2008qa}. In general, they both need the enveloping algebra of symmetries to be enlarged, including at least the Weyl generator, yielding for example
the deformed universal enveloping algebra of the Weyl group~\cite{kosinski1}.

\section{The bicrossproduct structure}\label{appB}
Following~\cite{Zaugg, agostini} we present here a constructive definition of the  bicrossproduct structure.

Let $\mathcal{X},\mathcal{A}$ be two Hopf algebras, a \textit{bicrossproduct algebra} $\mathcal{X} \vartriangleright \! \blacktriangleleft \mathcal{A}$ is the tensor product $\mathcal{X} \otimes \mathcal{A}$ endowed with two additional structure maps, a covariant right action of  $\mathcal{X}$ on $\mathcal{A}$ and a covariant left coaction of $\mathcal{A}$ on $\mathcal{X}$
\begin{subequations}
\begin{align}
\triangleleft &: \mathcal{A}\times \mathcal{X}\rightarrow \mathcal{A},\\
\beta &: \mathcal{X}\rightarrow \mathcal{A}\otimes \mathcal{X},
\end{align}
\end{subequations}
such that
\begin{subequations}\label{coac}
    \begin{align}
        a\triangleleft (xy) &=(a\triangleleft x)\triangleleft y,\\
        1\triangleleft x &=\varepsilon(x)1,\\
        (a\cdot b)\triangleleft x&=(a\triangleleft x_{(1)})(b\triangleleft x_{(2)}),\\
        (id \otimes \beta)\circ \beta &=(\Delta \otimes id) \circ \beta,\\
        (\varepsilon \otimes id) \circ \beta &= id,\\
        \beta(ab) &=\beta (a)\beta(b),\\
        \beta(1) &= 1\otimes 1,
    \end{align}
\end{subequations}
and with an Hopf algebra structure given by:
\begin{subequations}\label{hopfbicross}
\begin{align}
\mu ((x\otimes a),(y\otimes b)) &=(x\otimes a) \cdot (y\otimes b)=xy_{(1)}\otimes (a\triangleleft y_{(2)})b,\label{1.5a}\\
1_{\mathcal{X} \vartriangleright \! \blacktriangleleft \mathcal{A}} &=1_\mathcal{X}\otimes 1_\mathcal{A},\label{1.5b}\\
\Delta(x\otimes a) &=(x_{(1)}\otimes {x_{(2)}}^{(\bar{1})}a_{(1)})\otimes ({x_{(2)}}^{(\bar{2})}\otimes a_{(2)}),\label{1.5c}\\
\varepsilon (x\otimes a) &=\varepsilon (x)\varepsilon (a),\label{1.5d}\\
S(x\otimes a) &=(1_{\mathcal{X}} \otimes S(x^{(\bar{1})}a)) \cdot (S(x^{(\bar{2})})\otimes 1_{\mathcal{A}})\label{1.5e},
\end{align}
\end{subequations}
where $x,y \in \mathcal{X}$, $a,b\in \mathcal{A}$ and following the Sweedler notation, we have defined $\Delta (h)=\sum_{i} {h_{(1)}}_i \otimes {h_{(2)}}_i = h_{(1)} \otimes h_{(2)}$ and $\beta (x)=x^{(\bar{1})}\otimes x^{(\bar{2})}$, with $x^{(\bar{1})} \in \mathcal{A}$ and $x^{(\bar{2})} \in \mathcal{X}$.
The structure maps must also satisfy the following compatibility conditions:
\begin{subequations}\label{comp}
\begin{align}
\varepsilon (a\triangleleft x) &=\varepsilon (a)\varepsilon (x),\label{1.6a}\\
\Delta (a\triangleleft x) &=(a_{(1)}\triangleleft x_{(1)}){x_{(2)}}^{(\bar{1})}\otimes (a_{(2)}\triangleleft {x_{(2)}}^{(\bar{2})}),\label{1.6b}\\
\beta (xy) &=(x^{(\bar{1})}\triangleleft y_{(1)}){y_{(2)}}^{(\bar{1})}\otimes x^{(\bar{2})}{y_{(2)}}^{(\bar{2})},\label{1.6c}\\
{x_{(1)}}^{(\bar{1})}(a\triangleleft x_{(2)})\otimes {x_{(1)}}^{(\bar{2})} &=(a\triangleleft x_{(1)}){x_{(2)}}^{(\bar{1})}\otimes {x_{(2)}}^{(\bar{2})}\label{1.6d}.
\end{align}
\end{subequations}
An important thing to note is that the bicrossproduct algebra $\mathcal{X} \vartriangleright \! \blacktriangleleft \mathcal{A}$ can always be seen as the universal enveloping algebra generated by elements $X=x\otimes 1$, $\mathcal{A}=1\otimes a$, modulo the commutation relations
\begin{equation}
[X,A]=x\otimes a -x_{(1)}\otimes (a\triangleleft x_{(2)}); \label{1.7}
\end{equation}
in fact, from~\eqref{1.5a}
\begin{subequations}
\begin{align}
XA &=(x\otimes 1)\cdot (1\otimes a)=x\otimes (1\triangleleft 1)a=x\otimes a,\\
AX &=(1\otimes a) \cdot (x\otimes 1) =x_{(1)}\otimes (a\triangleleft x_{(2)}).
\end{align}
\end{subequations}

\subsection{The classical \texorpdfstring{$r$}{}-matrix deformation method}\label{appB1}
Here we give a short review of  the deformation  induced by the  \textit{classical $r$-matrix}, while referring  to~\cite{charipressley, ZakrzewskiInventsKPGroup, Lukierski_kappaPoincareanydimension} for details.

Given a Lie algebra $\mathfrak{g}$ corresponding to a Lie group $G$, a classical r-matrix is a tensor $r\in \bigwedge^2\mathfrak{g}$ satisfying the modified Yang-Baxter Equation (MYBE)
\begin{equation}\label{mybe}
    [r_{12},r_{13}+r_{23}]+[r_{13},r_{23}]=t,
\end{equation}
with $t\in \otimes^3\mathfrak{g}$ a $\mathfrak{g}$-invariant, and $r_{\alpha\beta}\in \otimes^3\mathfrak{g}$, $\alpha,\beta=1,2,3$, defined as
\begin{align}
    r_{12}=c_{ij}a_i\otimes a_j\otimes 1,\\
    r_{23}=c_{ij}1\otimes a_i\otimes a_j,\\
    r_{13}=c_{ij}a_i\otimes 1\otimes a_j,
\end{align}
$a_i\in \mathfrak{g}$.
If $t=0$ Eq.~\eqref{mybe} is called classical Yang-Baxter Equation (CYBE).

This element defines a Poisson-Lie group~\cite{charipressley} through the following \textit{Sklyanin bracket}:
\begin{equation}\label{poili}
    \{ f,g\}=r^{\alpha \beta}(X^R_\alpha f X^R_\beta g-X^L_\alpha f X^L_\beta g), \qquad f,g\in \mathcal{C}^\infty (G),
\end{equation}
where $X^R, X^L$ are the right- and left-invariant vector fields.
The algebra $\mathcal{C}^\infty (G)$ is, then, a Poisson-Hopf algebra with trivial cosector and antipodes.

If there is no order ambiguity, it is then possible to quantize this structure to a quantum group via the canonical quantization of the Sklyanin bracket $\{,\}\rightarrow \frac{1}{i}[,]$.

In the next section we shall apply~\eqn{poili} to the $\kappa$-Poincar\'e case, whereas in section \ref{sec2} it shall be applied to the $\varrho$-Poincar\'e deformation. Therefore we shall need    the left and right invariant vector fields of the Poincar\'e group explicitly. They read (see for example \cite{LSV})
\begin{equation}\label{lrfields}
\begin{array}{lll}
X_{\alpha \beta}^L &={\Lambda^\mu}_\alpha \frac{\partial}{\partial \Lambda^{\mu\beta}} -{\Lambda^\mu}_\beta \frac{\partial}{\partial \Lambda^{\mu\alpha}},  
& X_\alpha^L ={\Lambda^\mu}_\alpha \frac{\partial}{\partial a^\mu},\\
X_{\alpha \beta}^R &=\Lambda_{\beta\nu} \frac{\partial}{\partial {\Lambda^\alpha}_\nu} -\Lambda_{\alpha\nu} \frac{\partial}{\partial {\Lambda^\beta}_\nu}+a_\beta \frac{\partial}{\partial a^\alpha} -a_\alpha \frac{\partial}{\partial a^\beta},\;\;\;\;\;\;
& X_\alpha^R =\frac{\partial}{\partial a^\alpha}.
\end{array}
\ee
\subsection{The Quantum Group \texorpdfstring{$\mathcal{C}_\kappa (P)$}{}}\label{appB2}
In order to  obtain the quantum group $\mathcal{C}_\kappa (P)$, one  could  start from the $\kappa$-Minkowski commutation relations~\eqref{kappacomm}
and  impose the covariance of the $\kappa$-Minkowski algebra  under the action of the quantum group. However, as discussed in~\cite{LSV}, this method leads to some ambiguity in the mixed commutators between Lorentz and translation parameters. To avoid this issue, we will work in the framework of classical $r$-matrices.

A classical $r$-matrix for $\mathcal{C}_\kappa(P)$ is found to be~\cite{kosinski1}:
\begin{equation}
r=i\lambda M_{0\nu} \wedge P^{\nu}.
\end{equation}
This $r$-matrix satisfies a modified Yang-Baxter equation, which is why the twist operator requires an enlargement of the Poincar\'e algebra  as mentioned at the end of App.~\ref{appA}.

Following~\cite{LSV} we compute the Poisson brackets for the group-parameters using Eqs.~\eqn{poili} and~\eqn{lrfields} and quantize  them via  canonical quantization to obtain the following quantum group structure:
\begin{subequations}
\begin{align}
    [a^\mu, a^\nu]&=i\lambda ({\delta ^\mu}_0 a^\nu - {\delta ^\nu} _0 a^\mu),\\
    [{\Lambda^\alpha}_\beta,a^\rho] &=-i \lambda (({\Lambda^\alpha}_0 -{\delta^\alpha}_0 ){\Lambda^\rho}_\beta +(\Lambda_{0\beta}-g_{0\beta})g^{\alpha \rho}),\\
    [{\Lambda^\alpha}_\beta,{\Lambda^\gamma}_\delta] &=0,\\
    \Delta (a^\mu)&={\Lambda^\mu}_\nu \otimes a^\nu+ a^\mu \otimes 1,\\
\Delta ({\Lambda^\mu}_\nu)&= {\Lambda^\mu}_\alpha \otimes {\Lambda^\alpha}_\nu,\\
\varepsilon (a^\mu)&=0,\\
\varepsilon ({\Lambda^\mu}_\nu)&={\delta^\mu}_\nu,\\
S(a^\mu) &=-a^\nu {(\Lambda^{-1})^\mu}_\nu,\\
S({\Lambda^\mu}_\nu) &={(\Lambda^{-1})^\mu}_\nu.
\end{align}
\end{subequations}

This quantum group has a well-known bicrossproduct structure given in terms of the decomposition
\begin{equation}
\mathcal{C}_\kappa(P) =\mathcal{T}_\kappa^* \vartriangleright \! \blacktriangleleft \mathcal{C}(SO(1,3)),
\end{equation}
where $C(SO(1,3))$ is the classical commutative algebra of continuous functions on the Lorentz group and $\mathcal{T}_\kappa^*$ the algebra of functions on the dual of the translational sector, defined by
\begin{subequations}
\begin{align}
[a^\mu, a^\nu] &=i\lambda ({\delta ^\mu}_0 a^\nu - {\delta ^\nu} _0 a^\mu),\\
\Delta (a^\mu) &=a^\mu \otimes 1+1\otimes a^\mu,\\
S(a^\mu) &=-a^\mu,\\
\varepsilon(a^\mu) &=0.
\end{align}
\end{subequations}
The left coaction and the right action are respectively given by
\begin{subequations}
\begin{align}
\beta_L(x^\mu) &={\Lambda^\mu}_\nu \otimes x^\nu,\\
{\Lambda^\alpha}_\beta \vartriangleleft x^\varrho &= -{i}\lambda(({\Lambda^\alpha}_0 -{\delta^\alpha}_0 ){\Lambda^\rho}_\beta +(\Lambda_{0\beta}-g_{0\beta})g^{\alpha \rho}),
\end{align}
\end{subequations}
where the right action is read off   by the mixed commutators, and the left coaction is implicitly defined by the coproduct, as discussed in~\cite{Majid:1994cy}.
In this way we can explicitly see that $\mathcal{T}_\kappa^*\sim \mathcal{M}_\kappa$, and we can obtain the noncommutative spacetime from the quantum group via a quotient procedure similar to that of the classical case, noting that in this quantum version the quotient should be performed with respect to the deformed action.

Since this result is well-known in the literature, we will omit the details. We exhibit, in this paper, explicit calculations only  for the novel case of $\mathcal{C}_\varrho(P)$.
\section{Proof of Proposition \ref{Proprho}} \label{appproofcrho}
In order to complete the proof of Proposition \ref{Proprho}
 we have to verify that the action and the coaction satisfy~\eqref{1.6a}-\eqref{1.6d}.

We start proving~\eqref{1.6a}. From~\eqref{1.9b} and~\eqref{1.1g}, and since the counit map is a homomorphism:
\begin{equation}
\varepsilon ({\Lambda^\mu}_\nu \triangleleft x^\rho)=i \varrho \left[{\delta^\rho}_0 (g_{2\nu} {\delta^\mu}_1 -g_{1\nu} {\delta^\mu}_2) -{\delta^\rho}_0 ({\delta^\mu}_1 g_{2\nu} -{\delta^\mu}_2 g_{1\nu} )\right]=0,
\end{equation}
while, from~\eqref{1.1f},
\begin{equation}
\varepsilon({\Lambda^\mu}_\nu) \varepsilon ( x^\rho)=0,
\end{equation}
so that~\eqref{1.6a} is true.

We next consider~\eqref{1.6b}. Starting from the left-hand side we have that (omitting $i\varrho$ factors)
\begin{equation}
\Delta ({\Lambda^\alpha}_\beta \triangleleft x^\rho)=\Delta({\Lambda^\rho}_0)[\Delta ({\Lambda^\alpha}_1)g_{2\beta} -\Delta ({\Lambda^\alpha}_2)g_{1\beta}]-{\delta^\rho}_0 [\Delta (\Lambda_{2\beta}) {\delta^\alpha}_1- \Delta (\Lambda_{1\beta}) {\delta^\alpha}_2],
\end{equation}
since the coproduct is an homomorphism.

For the right-hand side we have
\begin{align}
&({\Lambda^\alpha}_\lambda \triangleleft x^\rho)\otimes {\Lambda^\lambda}_\beta +{\Lambda^\alpha}_\lambda {\Lambda^\rho}_\sigma \otimes ( {\Lambda^\lambda}_\beta \triangleleft x^\sigma)\nonumber \\
&={{\Lambda^\alpha}_\lambda {\Lambda^\rho}_\sigma \otimes {\Lambda^\sigma}_0 ({\Lambda^\lambda}_1g_{2\beta}- {\Lambda^\lambda}_2g_{1\beta})-\delta^\rho}_0 (\Lambda_{2\lambda} {\delta^\alpha}_1 -\Lambda_{1\lambda} {\delta^\alpha}_2) \otimes {\Lambda^\lambda}_\beta.
\end{align}
The terms factorized by ${\delta^\rho}_0$ are exactly equal to those coming from the coproducts factorized by ${\delta^\rho}_0$ in the left-hand side; moreover, the remaining terms can be checked to be equal to the remaining ones of the left-hand side, thus proving~\eqref{1.6b}.

For~\eqref{1.6c}, from the right-hand side we obtain
\begin{equation}
\beta ([x^\mu, x^\nu])=[{\delta^\mu}_0 (\Lambda_{2\alpha} {\delta^\nu}_1-\Lambda_{1\alpha} {\delta^\nu}_2)-{\delta^\nu}_0 (\Lambda_{2\alpha} {\delta^\mu}_1-\Lambda_{1\alpha} {\delta^\mu}_2)]\otimes x^\alpha, \label{1.17}
\end{equation}
while the right-hand side is
\begin{align}
\beta ([x^\mu, x^\nu]) =&({\Lambda^\mu}_\alpha \triangleleft x^\nu -{\Lambda^\nu}_\alpha \triangleleft x^\mu)\otimes x^\alpha+{\Lambda^\mu}_\alpha {\Lambda^\nu}_\beta \otimes [x^\alpha, x^\beta]\nonumber \\
=& [{\Lambda^\nu}_0 ({\Lambda^\mu}_1 g_{2\alpha} -{\Lambda^\mu}_2 g_{1\alpha} )-{\delta^\nu}_0 (\Lambda_{2\alpha} {\delta^\mu}_1 -\Lambda_{1\alpha} {\delta^\mu}_2) \nonumber\\
&+{\delta^\mu}_0 (\Lambda_{2\alpha} {\delta^\nu}_1 +\Lambda_{1\alpha} {\delta^\nu}_2) -{\Lambda^\mu}_0 ({\Lambda^\nu}_1 g_{2\alpha} -{\Lambda^\nu}_2 g_{1\alpha} )]\otimes x^\alpha\nonumber\\
&+{\Lambda^\mu}_\alpha {\Lambda^\nu}_\beta \otimes [{\delta^\alpha}_0 (x_2 {\delta^\beta}_1 -x_1 {\delta^\beta}_2)-{\delta^\beta}_0(x_2 {\delta^\alpha}_1-x_1 {\delta^\alpha}_2)];
\end{align}
terms factorized by ${\delta^\nu}_0$ and ${\delta^\mu}_0$ are equal to~\eqref{1.17}, and the remaining ones cancel out.

Let us turn our attention to~\eqref{1.6d}. To perform the calculation we recall~\eqref{1.3b} and~\eqref{1.9a}. Computing the left-hand side we have:
\begin{equation}
{x^\nu_{(1)}}^{(\bar{1})}({\Lambda^\alpha}_\beta\triangleleft x^\nu_{(2)})\otimes {x^\nu_{(1)}}^{(\bar{2})} = {\Lambda^\nu}_\lambda ({\Lambda^\alpha}_\beta \triangleleft 1) \otimes x^\lambda +({\Lambda^\alpha}_\beta \triangleleft x^\nu) \otimes 1,
\end{equation}
while, for the right-hand side:
\begin{equation}
({\Lambda^\alpha}_\beta\triangleleft x^\nu_{(1)}){x^\nu_{(2)}}^{(\bar{1})}\otimes {x^\nu_{(2)}}^{(\bar{2})}=({\Lambda^\alpha}_\beta \triangleleft 1) {\Lambda^\nu}_\lambda \otimes x^\lambda +({\Lambda^\alpha}_\beta \triangleleft x^\nu) \otimes 1,
\end{equation}
and by means of the commutativity of $\Lambda$'s, the compatibility request is proved.

\section{Some explicit calculations of the commutation relations for $\mathcal{U}_\varrho(\mathfrak{p})$
}\label{append}
In this appendix we provide explicit details of the computations needed to prove that the algebra sector of the novel $\varrho$-Poincaré basis remains undeformed. 

We refer to the change of basis~\eqref{redef} and remind the reader that the un-tilded generators close the standard Poincaré algebra. We want to prove that, after the change of basis, the algebra sector remains the same. 

The commutators involving only rotations and momenta are easily retrieved.  As an example, let us compute 
\begin{equation}
    \begin{aligned}
        [\tilde{R_1},\tilde{P_2}]=&\left[R_1\cos(\frac{\varrho}{2}P_0)-R_2\sin(\frac{\varrho}{2}P_0),P_2\cos(\frac{\varrho}{2}P_0)+P_1\sin(\frac{\varrho}{2}P_0)\right]\\
        =&iP_3\cos^2(\frac{\varrho}{2}P_0)+iP_3\sin^2(\frac{\varrho}{2}P_0)=iP_3=i\tilde{P_3}.
\end{aligned}
\end{equation}
When the  the third component of momentum is involved we have,
\begin{equation}
    \begin{aligned}
    [\tilde{R_1},\tilde{P_3}]=&[R_1\cos(\frac{\varrho}{2}P_0)-R_2\sin(\frac{\varrho}{2}P_0),P_3]\\
    =&-iP_2\cos(\frac{\varrho}{2}P_0)-iP_1\sin(\frac{\varrho}{2}P_0)=-i\tilde{P_2},
\end{aligned}
\end{equation}
giving the standard Poincaré result. With analogous  calculations we conclude that
\be
[\tilde{R_i},\tilde{P_j}]= i  \varepsilon_{ij}{}^k P_k.
\ee
Lie brackets involving the boost generators are more involved, since they imply an action of the boost generators  on the trigonometric functions. 
Let us compute, as an example,  
\begin{equation}
\label{exboostmom}
    \begin{aligned}
    [\tilde{N_1},\tilde{P_1}]=\left[N_1\cos(\frac{\varrho}{2}P_0)-N_2\sin(\frac{\varrho}{2}P_0)+\frac{\varrho}{2}R_3 \tilde{P_1},P_1\cos(\frac{\varrho}{2}P_0)-P_2\sin(\frac{\varrho}{2}P_0)\right].
\end{aligned}
\end{equation}
By exploiting the following result (easily proven by induction) 
\begin{equation*}
    [N_1,(P_0)^k]=ik(P_0)^{k-1}P_1, \qquad k\in\mathbb{N},
\end{equation*}
the commutators involving boosts and trigonometric functions of $P_0$ can be verified to be
\begin{equation}
    \begin{aligned}
        [N_1,\cos(\frac{\varrho}{2}P_0)]=-i\frac{\varrho}{2}\sin(\frac{\varrho}{2}P_0)P_1, \qquad [N_1,\sin(\frac{\varrho}{2}P_0)]=i\frac{\varrho}{2}\cos(\frac{\varrho}{2}P_0)P_1.
    \end{aligned}
\end{equation}
An analogous calculation is also valid for $N_2$.
Threefore we can compute all non-trivial terms in~\eqref{exboostmom}.
\begin{equation}
    \begin{aligned}
    &[N_1\cos(\frac{\varrho}{2}P_0),P_1\cos(\frac{\varrho}{2}P_0)]=iP_0\cos^2(\frac{\varrho}{2}P_0)-i\frac{\varrho}{2}P_1^2\cos(\frac{\varrho}{2}P_0)\sin(\frac{\varrho}{2}P_0),\\
    &[N_1\cos(\frac{\varrho}{2}P_0),P_2\sin(\frac{\varrho}{2}P_0)]=i\frac{\varrho}{2}P_1P_2\cos^2(\frac{\varrho}{2}P_0),\\
    &[N_2\sin(\frac{\varrho}{2}P_0),P_1\cos(\frac{\varrho}{2}P_0)]=-i\frac{\varrho}{2}P_1P_2\sin^2(\frac{\varrho}{2}P_0),\\
    &[N_2\sin(\frac{\varrho}{2}P_0),P_2\sin(\frac{\varrho}{2}P_0)]=iP_0\cos^2(\frac{\varrho}{2}P_0)+i\frac{\varrho}{2}P_2^2\cos(\frac{\varrho}{2}P_0)\sin(\frac{\varrho}{2}P_0).
    \end{aligned}
\end{equation}
Putting these results  together with the correct signs displayed in~\eqref{exboostmom} we obtain 
\begin{equation}
    i(P_0-\frac{\varrho}{2}(P_1\cos(\frac{\varrho}{2}P_0)-P_2\sin(\frac{\varrho}{2}P_0))(P_2\cos(\frac{\varrho}{2}P_0)+P_1\sin(\frac{\varrho}{2}P_0)))=i(\tilde{P_0}-\frac{\varrho}{2}\tilde{P_1}\tilde{P_2}).
\end{equation}
The remaining term in the commutator involving the rotation generator yields
\begin{equation}
    [\frac{\varrho}{2}\tilde{P_1}R_3,\tilde{P_1}]=i\frac{\varrho}{2}\tilde{P_1}\tilde{P_2},
\end{equation}
exactly canceling the other term proportional to $\frac{\varrho}{2}$. Therefore we conclude that
\begin{equation}
    [\tilde{N_1},\tilde{P_1}]=i\tilde{P_0}.
\end{equation}
Similar techniques can  be used to show that the commutators between the novel boost generators and the novel translation generators are the standard Poincaré ones. 

As for the Lorentz sector, analogous computations show that the commutators between rotations and boosts are also undeformed. The tricky ones are those involving only boost generators. Let us analyze the one between $\tilde{N_1}$ and $\tilde{N_2}$:
\begin{equation}
\begin{aligned}
    \label{n1n2}
    &[\tilde{N_1},\tilde{N_2}]=\\
    &=[N_1\cos(\frac{\varrho}{2}P_0)-N_2\sin(\frac{\varrho}{2}P_0)+\frac{\varrho}{2}R_3 \tilde{P_1},N_2\cos(\frac{\varrho}{2}P_0)+N_1\sin(\frac{\varrho}{2}P_0)+\frac{\varrho}{2}R_3 \tilde{P_2}].
\end{aligned}
\end{equation}
Introducing the shorthand notation 
\begin{equation}
    \begin{aligned}
   &A=N_1\cos(\frac{\varrho}{2}P_0)-N_2\sin(\frac{\varrho}{2}P_0), \qquad B=\frac{\varrho}{2}R_3 \tilde{P_1},\\
   &C=N_2\cos(\frac{\varrho}{2}P_0)+N_1\sin(\frac{\varrho}{2}P_0), \qquad D=\frac{\varrho}{2}R_3 \tilde{P_2},
    \end{aligned}
\end{equation}
we need to compute the quantity $[A,C]+[B,C]+[A,D]+[B,D]$. We start with $[A,C]$ and further break it down in commutators involving the  boosts with no tilde.
\begin{equation}
    \begin{aligned}
    &[N_1\cos(\frac{\varrho}{2}P_0),N_2\cos(\frac{\varrho}{2}P_0)]=[N_1,N_2]\cos(\frac{\varrho}{2}P_0)+N_1[\cos(\frac{\varrho}{2}P_0),N_2]\cos(\frac{\varrho}{2}P_0)\\
    &+N_2[N_1,\cos(\frac{\varrho}{2}P_0)]\cos(\frac{\varrho}{2}P_0)=i\bigl(-R_3\cos^2(\frac{\varrho}{2}P_0)+\frac{\varrho}{2}\cos(\frac{\varrho}{2}P_0)\sin(\frac{\varrho}{2}P_0)(N_1P_2-N_2P_1)\bigr),\\
    &[N_1\cos(\frac{\varrho}{2}P_0),N_1\sin(\frac{\varrho}{2}P_0)]=i\frac{\varrho}{2}N_1P_1,\\
    &[N_2\sin(\frac{\varrho}{2}P_0),N_2\cos(\frac{\varrho}{2}P_0)]=-i\frac{\varrho}{2}N_2P_2,\\
    &[N_2\sin(\frac{\varrho}{2}P_0),N_1\sin(\frac{\varrho}{2}P_0)]=i\bigl(R_3\sin^2(\frac{\varrho}{2}P_0)+\frac{\varrho}{2}\sin(\frac{\varrho}{2}P_0)\cos(\frac{\varrho}{2}P_0)(N_1P_2-N_2P_1).
    \end{aligned}
\end{equation}
Adding these up with the signs dictated by~\eqref{n1n2}, we get
\begin{equation}
    [A,C]=i\bigl(-R_3+\frac{\varrho}{2}{N_1P_1}+\frac{\varrho}{2}{N_2P_2}\bigr).
\end{equation}
Next, we compute $[A+B,D]$, given by
\begin{equation}
    [\tilde{N_1},\frac{\varrho}{2}R_3\tilde{P_2}]=\frac{\varrho}{2}[\tilde{N_1},R_3]\tilde{P_2}+\frac{\varrho}{2}[\tilde{N_1},\tilde{P_2}]=-i\frac{\varrho}{2}\tilde{N_2}\tilde{P_2}.
\end{equation}
Then, $[B,C]$ can also be easily calculated
\begin{equation}
\begin{aligned}
    [B,C]=[\frac{\varrho}{2}R_3\tilde{P_1},\tilde{N_2}-\frac{\varrho}{2}R_3\tilde{P_2}]=i\bigl(-\frac{\varrho}{2}\tilde{N_1}\tilde{P_1}+\frac{\varrho^2}{4}R_3\tilde{P_1}^2+\frac{\varrho^2}{4}R_3\tilde{P_2}^2\bigr).
\end{aligned}
\end{equation}
Collecting the results we have
\begin{equation}
    [\tilde{N_1},\tilde{N_2}]=i\bigl(-R_3+\frac{\varrho}{2}N_1P_1+\frac{\varrho}{2}N_2P_2-\frac{\varrho}{2}\tilde{N_2}\tilde{P_2}-\frac{\varrho}{2}\tilde{N_1}\tilde{P_1}+\frac{\varrho^2}{4}R_3\tilde{P_1}^2+\frac{\varrho^2}{4}R_3\tilde{P_2}^2\bigr).
\end{equation}
Using~\eqref{redef}, we find 
\begin{equation}    \tilde{N_1}\tilde{P_1}+\tilde{N_2}\tilde{P_2}=N_1P_1+N_2P_2+\frac{\varrho}{2}R_3\tilde{P_2}^2++\frac{\varrho}{2}R_3\tilde{P_1}^2,
\end{equation}
so that
\begin{equation}
    [\tilde{N_1},\tilde{N_2}]=-iR_3=-i\tilde{R_3},
\end{equation}
 like the in the standard Poincaré case. Commutators between $\tilde{N_1},\tilde{N_2}$ with $\tilde{N_3}$ are computed in a similar fashion, but with fewer intermediate steps.
\section{Plane waves in \texorpdfstring{$\kappa$}{}-Minkowski}\label{appe}
The $\kappa$-Minkowski spacetime is characterized by a noncommutativity between coordinates given in~\eqref{kappacomm}.
Following \cite{Freidel:2007hk,Kowalski-Glikman:2013rxa}, we  represent the coordinates in terms of $5\times 5$ matrices as follows:
\begin{equation}
\label{kcorrrep}
    \hat x^0=-i\ell \begin{pmatrix}
        0 & 0 & 0 & 0 & 1 \\
        0 & 0 & 0 & 0 & 0  \\
        0 & 0 & 0 & 0 & 0 \\
        1 & 0 & 0 & 0 & 0 
    \end{pmatrix} , \quad 
      \hat x^i=i\ell \begin{pmatrix}
        0 & -\mathbf{e}^i & 0\\
        -{\mathbf{e}^i}^T & \mathbf{0}_{3\cross 3} & -{\mathbf{e}^i}^T  \\
        0 & \mathbf{e}^i & 0 
    \end{pmatrix},
\end{equation}
where $\{ \mathbf{e}^i \}_{i=1,2,3}$ is the canonical vector basis for $\mathbb{R}^3$. 

We find that also in this case the time-to-the-right ordering of plane waves is linked to the translation generators in the bicrossproduct basis.
Indeed, noncommutative $\kappa$-plane waves in the time-to-the-right ordering, 
\begin{equation}
    \hat\phi_R(p)=e^{ip_i\hat x^i}e^{ip_0\hat x^0},
\end{equation}
lead to the composition law of momenta 
\begin{equation}
\begin{cases}
        (p\oplus_R k)_0=p_0+k_0,\\
        (p\oplus_R k)_i=p_i+e^{-\ell p_0}k_i.
\end{cases} 
\end{equation}
Defining the time-to-the-right translation operator $P^R_\mu$ such that $\hat \phi(p\oplus_R k)$ be an eigenfunction,
\begin{equation}
    P^R_\mu \hat\phi(p\oplus_R k)=(p\oplus_R k)_\mu  \hat\phi(p\oplus_R k),
\end{equation}
one can show that the coproducts of the time-to-the-right translation generators are the well-known bicrossproduct ones 
\begin{equation}
\begin{aligned}
   & \Delta P^R_0=P^R_0\otimes 1 + 1\otimes P^R_0,  \\
    &\Delta P^R_i=P^R_i\otimes 1+e^{-\ell P^R_0}\otimes P_i^R\,.
\end{aligned}
\end{equation}
On the other hand,  noncommutative $\kappa$-plane waves in the time-symmetric ordering 
\begin{equation}
    \hat\phi_S(p)=e^{\frac{ip_0\hat x^0}{2}}e^{ip_i\hat x^i}e^{\frac{ip_0\hat x^0}{2}},
\end{equation}
lead to  the composition law
\begin{equation}
\begin{cases}
        (p\oplus_S k)_0=p_0+k_0,\\
        (p\oplus_S k)_i=p_ie^{\frac{\ell k_0}{2}}+e^{\frac{-\ell p_0}{2}}k_i\,.
\end{cases} 
\end{equation}
This is compatible with the coproducts for the time-symmetric translation generators, 
\begin{equation}
    P^S_\mu \hat\phi(p\oplus_S k)=(p\oplus_S k)_\mu \hat\phi (p\oplus_S k)\,,
\end{equation} 
of the form
\begin{equation}
\begin{aligned}
    &\Delta P^S_0=P^S_0\otimes 1 + 1\otimes 1\otimes P^S_0,\\
    & \Delta P_i^S= P_i^S \otimes e^{\frac{\ell  P^S_0}{2}}+ e^{\frac{{-\ell P^S_0}}{2}}\otimes P_i^S.
    \end{aligned}
\end{equation}
These are the coproducts of the $\kappa$-Poincaré translations in the so-called \textit{standard basis}~\cite{Lukierski:1992dt} and are related to the bicrossproduct ones via the map \cite{{Majid:1994cy}}
\begin{equation}
\label{kchange}
\begin{aligned}
       & P_0^R=P_0^S,\\
       & P_i^R=P_i^Se^{-\frac{\ell P_0^S}{2}}.
\end{aligned}
\end{equation}

\providecommand{\href}[2]{#2}\begingroup\raggedright\endgroup

\end{document}